\documentclass[aps,twocolumn,nofootinbib,preprintnumbers]{revtex4-1}
\usepackage{bm}
\usepackage{latexsym}
\usepackage{dcolumn}
\usepackage{amsfonts,amssymb}
\usepackage{graphicx,epsfig}
\usepackage{amsmath,amssymb}
\newcommand{\ben}{\begin{eqnarray}\displaystyle}
\newcommand{\een}{\end{eqnarray}}
\newcommand{\be}{\begin{equation}}
\newcommand{\ee}{\end{equation}}
\newcommand{\lb}{\left (}
\newcommand{\rb}{\right )}
\newcommand{\ltb}{\left [}
\newcommand{\rtb}{\right ]}

\newcommand{\ra}{\rightarrow}

\newcommand{\nn}{\nonumber}
\newcommand{\nt}{{1\over 16 \pi G_5}}
\newcommand{\app}{\alpha'}

\newcommand{\hhp}{\hspace{.15cm}}

\newcommand{\ep}{\epsilon}

\newcommand{\cA}{{\cal A}}

\newcommand{\ph}{\phi(r,k)}

\newcommand{\intk}{\int {d^4 k \over (2 \pi)^4}}

\newcommand{\higho}{{\cal O}(q\omega^2, \omega q^2, q^3, \omega^3)}

\newcommand{\sectiono}[1]{\section{#1}\setcounter{equation}{0}}

\def\lfig#1#2#3#4{
 \begin{figure}[h]
 \refstepcounter{figure}
 \label{#4}
 \addtocounter{figure}{-1}
 \epsfxsize=#3
 \centerline{\epsfbox{#2}}
 {\bf \caption{{\rm #1}}}
 \end{figure}
}

\begin{document}

\title{Nonlinear Hydrodynamics from Flow of Retarded Green's Function}
\author{Nabamita Banerjee}
\email[]{E-mail: N.Banerjee@uu,nl}
\affiliation{ITF, Utrecht, The Netherlands }
\author{Suvankar Dutta}
\email[]{E-mail: pysd@swan.ac.uk}
\affiliation{Dept. of Physics, Swansea University, UK}
%\date{\today}

%%%%%%%%%%%%%%%%%%%%%%%%%%%ABSTRACT%%%%%%%%%%%%%%%%%%%%%%%%%%%%%%%%%%%%%%%%%%

\begin{abstract}
  We study the radial flow of retarded Green's function of
  energy-momentum tensor and $R$-current of dual gauge theory
 in presence of generic higher derivative terms in bulk Lagrangian.
  These are first order non-linear $Riccati$ equations.
  We solve these flow equations
  analytically and obtain second order transport coefficients
of boundary plasma.
This
  way of computing transport coefficients
has an advantage over usual Kubo approach. The non-linear
equation turns out to be a linear first order equation when
we study the Green's function perturbatively
in momentum.
We consider several examples including
 $Weyl^4$ term and generic four derivative terms in bulk. We
 also study the flow equations for $R$-charged black
 holes and obtain exact expressions for second order transport
 coefficients for dual plasma in presence of arbitrary
 chemical potentials. Finally we obtain higher derivative
 corrections to second order transport coefficients of
 boundary theory dual to five dimensional gauge supergravity.
\end{abstract}

%\pacs{04.20.Jb, 04.62.+v, 98.80.Cq}

\maketitle

\tableofcontents

%%%%%%%%%%%%%%%%%%%%%%%%%%%%%%%%%%%%%%%%%%%%%%%%%%%%%%%%%%%%%%%%%%%%%%%%%%%%
\section{Introduction and Discussion}
\label{intro}

Fluid/gravity correspondence has become an interesting aspect of current theoretical physics
research after beginning of $RHIC$ program in 2000. The experimental data implies that
the quark-gluon plasma (QGP) produced in $RHIC$  are in a new state called
 thermalized matter. The evolution of QGP and hadronic matter in this state can be described by
 hydrodynamics. The
temperature of the gas of quarks and gluons produced at RHIC is approximately 170MeV
which is very close to the confinement temperature of QCD. At this high temperature
they are not in the weakly coupled regime of QCD. This new phase of nuclear
matter is known as the the strongly coupled quark-gluon plasma (sQGP). Obviously to study their properties the usual
perturbation theory does not work. One needs different approach
to deal with these strongly coupled system.

The holographic hydrodynamics (or fluid/gravity correspondence)
is an important tool for understanding some properties of strongly
coupled CFTs in terms of the dual AdS black holes physics\footnote{QCD is approximately
conformal at sufficiently large energies.} \cite{son1}-\cite{her}.

Low frequency (long wavelength) fluctuations of any interacting quantum
field theory should be  described by hydrodynamics.
Low energy behavior of strongly coupled field theory (with gravity dual)
is governed by a weakly coupled black hole space-time in one higher
dimension. On the other hand in classical theory of gravity a black hole
can be viewed as a fictitious fluid membrane (with hydrodynamic characteristic)
living on the horizon of the black hole \cite{Price:1986yy}. Therefore from UV/IR point of view
it would be interesting to understand the precise relation between the
membrane fluid and the low frequency description of strongly coupled field
theory sitting at the boundary of $AdS$ black hole space-time.

One can read off the transport coefficients like shear-viscosity
coefficients, of boundary plasma from its retarded Green's function
of stress tensor
\be
G_{xy,xy}^R(k_{\mu})= -i \int dt dx e^{i k\cdot x}\langle\ltb T_{xy}(x),T_{xy}(0)\rtb \rangle \ .
\ee
We write the Green's function in powers of momentum and it is given by\footnote{Here we have
dropped the frequency independent part of the Green's function.},
\be\label{Gdef}
G^R_{xy,xy}(k_{\mu}) = - i  \eta \omega + \eta \tau_{\pi} \omega^2 - {\kappa \over 2}[(p - 2)\omega^2 + q^2]
+ {\cal O}(k^3),
\ee
where, $p$ is spatial dimension and $p \geq 3$ and
\ben
%P &:& pressure,\\
\eta &:& \rm{shear \ viscosity \ coefficient},\nn\\
\tau_{\pi} &:& \rm{relaxation \ time \ for \ shear \ viscous \ stress},\nn\\
\kappa &:& \rm {a \ new \ coefficient \ defined \ in\ Ref[9,10]} \nn
\een
and $\omega$ is frequency and $q$ is spatial momentum ($k=\{\omega,0,0,q\}$).

In \cite{liu} it was shown that, at low momentum ($k_{\mu}\ra 0$), the evolution of
the boundary Green's function is independent of the radial direction and hence,
it can be computed either at horizon or at boundary. Computing the Green's
function at black hole horizon one can show that the shear viscosity
coefficient, which is a first-order transport coefficient of the boundary plasma, can
be obtained from the low frequency characteristics of the membrane fluid.
In \cite{cai1,BD1,BD2}, this issue has
been generalized for gravity theory with generic higher-derivative interactions in
arbitrary backgrounds. It has  been proved that the membrane fluid does give
the correct shear-viscosity coefficient of the boundary plasma in arbitrary higher
derivative gravity and just the knowledge of near-horizon geometry of the dual AdS
black hole is enough for computing the shear-viscosity coefficient \cite{BD3}.

To specify the boundary
plasma completely one also needs to understand its higher-order transport coefficients. For
this one needs to move away from the low frequency ($k_\mu \ra 0$) limit. In this case, the Green's
functions flow non-trivially with the radial direction and the flow depends on full black
hole geometry. Although the boundary plasma and the membrane fluid
have same shear-viscosity coefficients, other transport coefficients can
certainly differ and it is not clear how the two are related. In Fig. \ref{fig1}
we plot the radial evolution of response function for
two derivative gravity\footnote{We define the response function $\bar \chi={-G_{xy,xy}^R(\omega,q)\over i \omega}$.}.
\lfig{Flow of Green's function from horizon to boundary for two derivative gravity.}
{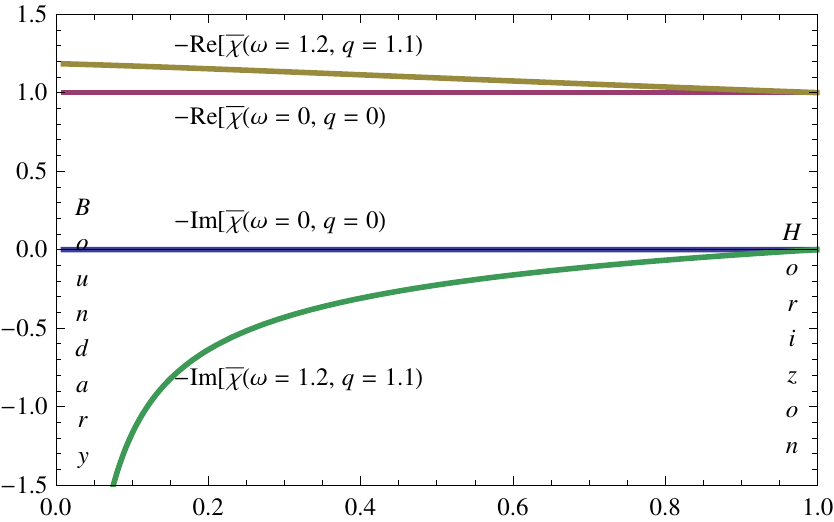}{6.5cm}{fig1}
In zero frequency limit flow of the response function $\bar{\chi}$ is trivial. Its
real part is constant (the constant is one in our scaling) and imaginary part
is zero. For finite $\omega$ and $q$ the response function (both real and imaginary part)
has a non-trivial evolution. But the horizon value of the function for non-zero frequency
 is same as the horizon value of the function for zero frequency. Therefore we conclude
 that for two derivative gravity dual, the full momentum response at
the horizon automatically corresponds to only the zero momentum limit of the boundary response.

However in presence of higher derivative terms in the action the situation is different.
The full momentum
response at horizon depends on spatial momentum\footnote{We will also see this
analytically in sections \ref{modelhd}.}. We plot the flow of response
function for $Weyl^4$ interaction in Fig. \ref{fig2}.

\lfig{Flow of Green's function from Horizon to boundary for higher derivative gravity for different values of $\omega$ and $q$.}
{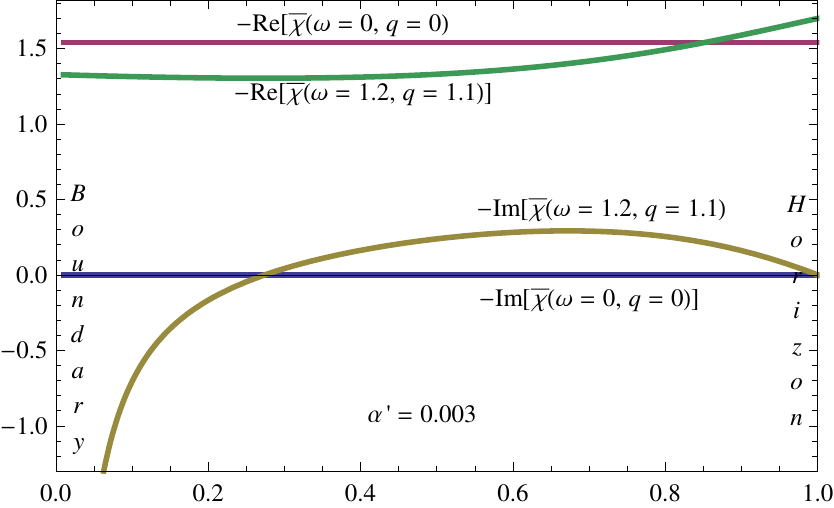}{6.5cm}{fig2}

From this plot we see that in presence of higher-derivative interaction
the horizon value of the response function depends of $k_{\mu}$
unlike two derivative theory.

We also see that the imaginary part of $\bar \chi$ diverges at
boundary. This is usual $UV$ divergence. One
has to add a counter term to cancel this divergence.

In this paper, we study the flow equations for retarded Green's function of boundary theory
analytically and find higher order transport coefficients of the boundary plasma solving this equation.
We generalize the analysis for generic higher derivative
gravity theory and also for $R$ charge black holes. The flow
equation for Green's function is a first order non-linear differential
equation of $Riccati$ type. Because of its non-linear nature it is hard
to solve this equation exactly. After a change of variable one can reduce this
non-linear equation to a second order linear homogeneous differential equation.
But to solve this we need to specify two boundary conditions.
In this paper we deal with the non-linear equation and specify the
boundary condition at the horizon. Therefore the hydrodynamic characteristic of the
field theory at $UV$ fix point is determined by $IR$ boundary condition.

For two derivative Einstein-Hilbert action the flow equation of retarded Green's function
has been derived in \cite{liu}.
But it is not obvious how to generalize the flow equation for higher derivative gravity.
The derivation given in \cite{liu}
was based on the canonical form of graviton's action. In this paper we have considered generic
higher derivatives terms in the bulk Lagrangian.
Following the prescription given in \cite{BD1} we construct an effective action for transverse graviton
which has the canonical form in presence of any higher derivative terms in the bulk,
and derive the flow equation for Green's function. Solving this flow equation perturbatively
in $\omega$ and $q$ we obtain second order transport coefficients namely $\tau_{\pi}$ (relaxation
time) and $\kappa$ of the dual plasma\footnote{From Weyl invariance one can show that there are other transport coefficients in second order
hydrodynamics \cite{Baier}. However from the expansion of retarded Green's function it is only possible
to compute only two of them.}. In this way of computing the transport coefficients
has an advantage over usual Kubo approach. In Kubo approach, one has to first find the transverse
graviton by solving a second order differential equation and then compute regarded Green's function.
Instead, the flow equation is a first order differential equation (although non-linear).
As we want a perturbative expansion of Green's function in powers of $\omega$ and $q$
 the equation turns out to be a linear first order differential equation. Thus, technically,
it is simpler to get results for causal hydrodynamics, particularly
when the dual bulk theory is complicated.

The paper is organized as follows. We have worked in five-dimensional bulk theory,
the dual gauge theory is four-dimensional. In section \ref{model} we review
the derivation of flow equation of boundary Green's function for two derivative gravity. Then
solving this flow equation we
compute second order transport coefficients $\kappa$ and $\tau_{\pi}$ of boundary plasma. Our results matches with previous
computations of \cite{shiraz1,Baier}.
In section \ref{modelhd} we present the flow equation
in presence of generic higher derivative interaction. We calculate the higher
derivative correction for $\tau_{\pi}$
and $\kappa$ in section \ref{hdcsec}. We concentrated mainly on
$Weyl^4$ and four derivative correction.  While our results for $Weyl^4$
correction are in agreement with results of \cite{buchel-paulos1}, the
four-derivative corrections are new results of this paper. We also
compute the effect of gauss-Bonnet to $\tau_{\pi}$ and $\kappa$. From our
results we obtain the bound on Gauss-Bonnet coupling constant. In section \ref{cbh}
we consider $R$ charge black hole in bulk and
find the second order transport coefficients in presence of finite chemical potential. We
also find the higher derivative effects on transport coefficients of field theory plasma
dual to gauge supergravity theory.
%in presence
%of small chemical potential and discuss its phenomenological aspects.
Finally,we analyze the flow equation of
Green's function for boundary $R$ current in section \ref{rcflow}. Appendix \ref{kubo} and \ref{BOUN}
compliments some discussion on boundary terms. In appendix \ref{weylapp}, \ref{gbapp} and \ref{hdcb}, we have provided some long
expressions.

%%%%%%%%%%%%%%%%%%%%%%%%%%%%%%%%%%%%%%%%%%%%%%%%%%%%%%%%%%%%%%%%%%%%%%%%%%%%%%%%%%%%%%%%%%%%
%%%%%%%%%%%%%%%%%%%%%%%%%%%%%%%%%%%%%%%%%%%%%%%%%%%%%%%%%%%%%%%%%%%%%%%%%%%%%%%%%%%%%%%%%%%%

\sectiono{Flow of retarded Green's function of
 energy-momentum tensor} \label{model}

In this section we briefly review the work of Liu and
Iqbal \cite{liu}. They considered leading
Einstein-Hilbert (E-H) action with a negative cosmological constant in 4+1
dimensions and studied the motion of a transverse graviton
in this background\footnote{Here we restrict ourselves to five
space-time dimensions, but the discussions are quite generic and
can be extended to arbitrary dimensions.}. The action is,
\be
\label{action1}
S_{{\rm EM}} = { 1 \over 16 \pi G_5} \int d^5x \sqrt{-g} \left ( R + {12} \right ).
\ee
The background has a black-brane solution as,
\ben\label{sol}
dS^2&=&  g_{tt} dt^2+ g_{rr} dr^2+ g_{ij}dx^2dx^j, \nn \\
g_{tt}&=&-{1-r^2 \over r}, \,\,\,\,\ g_{rr}={1 \over 4 r^2(1-r^2)}\nn \\
g_{ij}&=& {1 \over r}\delta_{ij}.
\een
The solution is asymptotically $AdS$ and it has a boundary topology $R \times R^3$.
The horizon of the space-time is at $r \ra 1$ and asymptotic boundary is at $r \ra 0$.

We study the graviton's fluctuation in this background,
\be\label{petmet}
g_{xy}=g^{(0)}_{xy}+ h_{xy}(r,x)=g^{(0)}_{xy}[1+\ep \Phi(r,x)].
\ee
By plugging it in the action and keeping terms to order
$\epsilon^2$, we obtain the following effective
action for the perturbation
\be \label{ghdacnphi1}
S=\nt \int {d^4 k \over (2 \pi)^4} dr \sum_{p,q=0}^{2} \cA_{p,q}(r,k)
\phi^{(p)}(r,-k) \phi^{(q)}(r,k).
\ee
Here we use Fourier transform to work in the momentum space $k=\{-\omega,\vec{k}\}$,
\be \label{phifu}
\Phi(r,x) = \int {d^4k \over (2 \pi)^4} e^{i k.x} \phi(r,k)
\ee
and $\phi^{(p)}(r,k)$ denotes the $p^{th}$ derivative of the field $\ph$
with respect to $r$ ($p+q\leq 2$).\\

Next, we integrate by parts to obtain the bulk action in the following form
(up to some total derivative terms)
\ben \label{ghdacnphi}
S&=& \int {d^4 k \over (2 \pi)^4} dr ({\cal A}_1(r,k) \phi'(r,k)\phi'(r,-k) \nn \\
&& \qquad \qquad +{\cal A}_0(r,k) \phi(r,k)\phi(r,-k) ),
\een
where,
\ben\label{coeff}
{\cal A}_1(r,k)&=&-{{1\over 2}g^{rr}\sqrt{-g}\over 16 \pi G_5}, \nn \\
{\cal A}_0(r,k)&=&- {{1 \over 2}\sqrt{-g}g^{\mu\nu}k_{\mu}k_{\nu}\over 16 \pi G_5}.
\een

From this action, we can find the conjugate momentum
$\Pi(r,k_\mu)$ of the transverse graviton (for r-foliation) and the equation of motion,
\be
\Pi(r,k_\mu)= 2 {\cal A}_1(r,k) \phi'(r,k)
\ee
and
\be\label{eom}
\Pi'(r,k_{\mu})-2\ {\cal A}_0(r,k) \phi(r,k)=0 \ .
\ee

The on-shell action reduces to the following surface term\footnote{We will discuss about other boundary terms in appendix \ref{BOUN}},
\be\label{boundacn}
S= \sum_{r=0,1}\int {d^4 k \over (2 \pi)^4 }({\cal A}_1(r,k) \phi'(r,k)\phi(r,-k)).
\ee
Following the AdS/CFT prescription given in \cite{son4}, the boundary retarded Green's function is given as,
\be
G_R(k_\mu)=\lim_{r \rightarrow 0}\frac{2{\cal A}_1(r,k) \phi'(r,k)\phi(r,-k)}{\phi_0(k) \phi_0(-k)},
\ee
where, $\phi_0(k_\mu)$ is the value of the graviton fluctuation at boundary.
% which in the dual prescription acts as a source to the boundary operator we are interested in.
Full solution of the graviton can be written as $\phi(r,k_\mu)= \phi_0(k_\mu)F(r,k_\mu)$, where $F(r,k_\mu)$
goes to identity at the boundary.
We can rewrite the boundary retarded Green's function as,
\ben
G_R(k_\mu)&=& \lim_{r \rightarrow 0}{\Pi(r,k_\mu)\over \phi(r,k_\mu)}.
\een
Let us define a response function of the boundary theory as\footnote{We set the zero frequency part of G to zero, as it gives contact terms},
\be\label{barchi}
\bar \chi(k_\mu,r)= {\Pi(r,k_\mu)\over  i \omega \phi(r,k_\mu)}
\ee
where $\omega=k_{0}$.
This function is defined for all $r$ and $k_\mu$. Therefor the boundary Green's
function is given by,
\be
G_R(k_\mu)=\lim_{r \rightarrow 0} i \omega \bar \chi(k_\mu,r).
\ee

We will study the radial evolution of the response function $\bar \chi(k_\mu)$
from horizon to boundary. Differentiating equation (\ref{barchi}) and using
the equations of motion (\ref{eom}) we get,
\be\label{flow}
\partial_r \bar \chi(k_{\mu},r)= i \omega \sqrt{- {g_{rr} \over g_{tt}}}
\Bigg[{\bar \chi(k_{\mu},r)^2 \over \Sigma(r)}- {\Upsilon(r) \over \omega^2}\Bigg],
\ee
where we define
\ben
\Sigma(r)&=& - 2 {\cal A}_1(r,k_\mu)\sqrt{-{g_{rr} \over g_{tt}}}\\
\Upsilon(r)&=& 2 {\cal A}_0(r,k_\mu)\sqrt{-{g_{tt} \over g_{rr}}}.
\een

Putting values of ${\cal A}_1$ and ${\cal A}_0$ given in
(\ref{coeff}) we can easily recover the
flow equation given in \cite{liu}\footnote{just notice
that $g_{tt}$ there is negative of what we used here.}.
However for future requirements, here we present it directly
in terms of the coefficients of the graviton action.

As mentioned earlier, the flow equation in (\ref{flow}) is
valid for any value of momentum. This is a first
order differential equation and we need to specify one
boundary condition to solve this equation. That naturally
comes from the behavior of the equation at the horizon.
 Demanding the solution to be regular at the horizon,
 we get the following condition,
\be\label{bocon}
\bar \chi({k_{\mu}},r)^2\Bigg|_{r=1}= {\Sigma(r) \Upsilon(r) \over \omega^2}\Bigg|_{r=1}.
\ee
For two derivative gravity this boundary condition implies that\footnote{We choose the negative brunch. The sign
of the boundary condition \ref{bocon} depends on the choice of coordinate. In our coordinate the boundary is at
$r\ra 0$ hence we need to choose the negative branch.},
\be\label{bocon2}
\bar \chi({k_{\mu}},1)=- \sqrt {{\Sigma(1) \Upsilon(1) \over \omega^2}}=- \nt
\ee
which is independent of $k_{\mu}$. Therefore the full momentum response at the
horizon corresponds to only to the zero momentum limit of
boundary response, $\bar\chi({k_{\mu}},1)=\bar\chi(k_{\mu}\ra 0,r \ra 0)$ \footnote{ However,
in higher derivative gravity we will see that the $\bar\chi(k_{\mu},1)$ depends
on spatial momentum.}.

With this boundary condition, one can integrate out the
differential equation (\ref{flow}) from horizon to
asymptotic boundary and obtain the AdS/CFT response
for all momentum $k_{\mu}$. In particular, it is
trivial to see that at $(\omega,k_{i}) \rightarrow 0$ limit,
the flow is trivial
\be
\partial_r \bar \chi(k_{\mu},r)=0
\ee
and using the boundary condition (\ref{bocon}) we get  the first order transport
coefficient of boundary fluid, i.e.
the shear viscosity coefficient coefficients turns out to be $\eta=\nt$.

In this paper, we will go away from $(\omega,k_i \rightarrow 0)$
limit. As we have already mentioned, it is possible to integrate
the flow equation for any momentum (perturbatively) and we can easily
find the higher order transport coefficients. The usual Kubo approach
to compute these coefficients requires the full profile of
the transverse graviton in black hole background background (solving a second order
differential equation), where as, using the
flow equation, one can get these transport coefficients without explicit knowledge of the
 graviton's profile.

%%%%%%%%%%%%%%%%%%%%%%%%%%%%%%%%%%%%%%%%%%%%%%%%%%%%%%%%%%%%%%%%%%%%%%%%%%%%%%%%
%%%%%%%%%%%%%%%%%%%%%%%%%%%%%%%%%%%%%%%%%%%%%%%%%%%%%%%%%%%%%%%%%%%%%%%%%%%%%%%%

\subsection{A renormalized response function}\label{renormalization}

When we solve the flow equation (\ref{flow}) to get the boundary response function
in general it involves divergence at the boundary ($r\ra 0$). These
are usual $UV$ divergences and to remove them we need to re-normalize
the response function properly.

We follow the holographic renormalization prescription of \cite{skende,skenda1}.
As the graviton is massless, we only need to add the following counterterm
to the graviton's action,
\be\label{CT}
S_{C}= {1\over 16 \pi G_5} \int_{r=\delta} d^4 x \sqrt{-\gamma} \frac{1}{4}\Phi(\epsilon, x)\Box\Phi(\epsilon, x).
\ee
In momentum space,
\be
S_{C}= {1\over 64 \pi G_5} \int_{r=\delta} {d^4 k\over (2\pi)^4} \sqrt{-\gamma} \phi(\delta, k)(g^{tt}\omega^2+k_ik^i)\phi(\delta, -k).
\ee
Therefore the renormalized Green's function is given by,
\be
{\cal G}_{R}= \lim_{r \rightarrow 0}\ltb{\Pi(r,k_\mu)\over \phi(r,k_\mu)}
+ \frac{\sqrt{-\gamma}}{32 \pi G_5} (g^{tt}\omega^2+k_ik^i)\rtb.
\ee

However we will study the flow of un-renormalized response function defined
in (\ref{barchi}) and we define our renormalized response function as,
\be
\bar{\chi}^{Ren}(r,k_{\mu}) =\bar{\chi}(r,k_{\mu})+\frac{1}{i \omega} { \sqrt{-\gamma} (g^{tt}\omega^2+k_ik^i)\over 32 \pi G_5}.
\ee
The counter term will cancel the UV divergences appearing
in the expression of $\bar{\chi}$ and we will get a finite result at the boundary, i.e.
$ \lim_{r \rightarrow 0}\bar{\chi}^{Ren}(r,k_{\mu})$ will be finite.
From the above analysis, we understand that one can get rid of the UV
divergences appearing in the response function by following the
holographic renormalization technique. But, an important
observation is, this counter term does not add any finite
contribution to the result it
only cancels out the divergences. Thus, one can study the flow
of the un-renormalized response function and
ignore the divergences piece to get the finite contribution
at the boundary.

%%%%%%%%%%%%%%%%%%%%%%%%%%%%%%%%%%%%%%%%%%%%%%%%%%%%%%%%%%%%%%%%%%%%%%%%%%%%%%%%
%%%%%%%%%%%%%%%%%%%%%%%%%%%%%%%%%%%%%%%%%%%%%%%%%%%%%%%%%%%%%%%%%%%%%%%%%%%%%%%%

\subsection{Second order transport coefficients from flow equation}

In this subsection, we compute the higher order transport coefficients by
solving the flow equation (\ref{flow}) perturbatively up to
order $\omega^2$ and $k_i^2$. This is a non-linear first order
differential equation. Now, the right hand side of this equation is
proportional to $\omega$. Hence, to solve $\bar {\chi}$ to order
$\omega^2$, we can replace the leading order
 solution for $\bar{\chi}$ in the right hand side of equation (\ref{flow}).
 This simplifies the situation a lot as the
 non-linear equation becomes linear.
 Now, to
leading order, $\bar{\chi}=-\eta=-{1 \over 16 \pi G_5}$.
Therefore up to order $\omega^2$, we get,
\be
\partial_r \bar \chi(k_\mu,r)= i \omega \sqrt{- {g_{rr} \over g_{tt}}}
\Bigg[{\eta^2 \over \Sigma(r)}- {\Upsilon(r) \over \omega^2}\Bigg] + {\cal O}(\omega^2,k_i^2).
\ee
The integration constant for the equation can be fixed form the boundary
condition (\ref{bocon}). Putting the value of the constant, the solution takes the form,
\ben\label{chi}
i \omega \bar \chi(k_{\mu},0)&=&\lim_{r\ra 0}-{1\over 96 \pi G_5 r}\bigg[3 q^2 (r-1)
+\omega  (3 \omega \nn\\
&& \qquad \qquad +r (\omega  (\log (8)-3)+6 i))\bigg]\nn\\
&& \qquad \qquad + \higho\nn\\
&=& -i\omega \lb \nt \rb  \nn\\
&& + \omega^2 \ltb {1\over 2}(1-\ln 2) \lb \nt \rb  \rtb\nn\\
&&  - {q^2 \over 2} \lb \nt \rb +{\cal O}({1\over r}).
\een
Here we have chosen the four momentum in the following form
\be
k= \{\omega,0,0,q\}\ .
\ee

This expression has divergence as $r\ra 0$ (UV divergence) and
can be removed by adding suitable counter term (as explained in the last section).

Comparing the finite piece of (\ref{chi}) at $r\ra 0$ with the
generic expansion of the retarded Green's function (\ref{Gdef})\footnote{The overall sign depends on
the choice of coordinate.},
we get,
\ben\label{htf}
\eta &=& {T^3 \pi^3\over 16 \pi G_5},\nn\\
\tau_{\pi}&=&{2-\ln2 \over 2 \pi T},\ \ \ \
 \kappa = {\eta\over \pi T}\ .
\een
Here $T={1\over \pi}$.
These results are in agreement with \cite{Baier}. In appendix \ref{kubo} we briefly
outline the Kubo method to get this result.
Thus, we see that,
studying the flow equation of the response function we can
compute the higher order transport coefficients perturbatively.
Here, we present the results for the second order transport
coefficients, but,
in general it is possible to go beyond second order.

At this point, it is not clear why only considering the
boundary term from action (\ref{ghdacnphi}) is enough to
get the correct results. In usual Kubo approach,
one needs to take into account the Gibbons-Hawking term also. But
as the action (\ref{ghdacnphi}) has well defined variational
principle, one does not need to add any Gibbons-Hawking term with it.
In appendix \ref{BOUN} we show that the boundary terms coming
from the original action and the corresponding
Gibbons-Hawking action are exactly same as the boundary
terms coming from the action (\ref{ghdacnphi}) up to terms
proportional to $\phi^2$ and pure divergence
terms. The $\phi^2$ terms do not contribute to any
transport coefficients\footnote{They only contribute to pressure of the boundary theory.}.
The divergent terms will get
canceled by the proper counterterms and hence are
not important for finding the transport coefficients.
Thus it is clear that the effective action will give
us the correct transport coefficients for the boundary plasma. This observation also holds
for higher derivative gravity theory\footnote{In appendix \ref{BOUN}, we have proved this explicitly
for $R^{(n)}$ gravity theory.}.

%%%%%%%%%%%%%%%%%%%%%%%%%%%%%%%%%%%%%%%%%%%%%%%%%%%%%%%%%%%%%%%%%%%%%
%%%%%%%%%%%%%%%%%%%%%%%%%%%%%%%%%%%%%%%%%%%%%%%%%%%%%%%%%%%%%%%%%%%%%
%%%%%%%%%%%%%%%%%%%%%%%%%%%%%%%%%%%%%%%%%%%%%%%%%%%%%%%%%%%%%%%%%%%%%%%

\sectiono{Higher derivative correction to flow equation}\label{modelhd}

So far we have discussed the flow equation of two point correlation
function of energy-momentum tensor of boundary theory whose gravity
dual is given by Einstein-Hilbert action (two derivative action).
But it is not obvious how to generalize this for higher derivative
case.  The proof given in \cite{liu} was based on the canonical form
(\ref{ghdacnphi}) of graviton's action.  In presence of arbitrary higher
derivative terms in the bulk, the general action for the perturbation $h_{xy}$
does not have the above form (\ref{ghdacnphi}). Rather it will have more than
two derivative (with respect to $r$) terms like $\phi'\phi''$, $\phi''^2$ $etc$.
In presence of these terms it is not possible to bring this action into a
canonical form (up to some total derivative terms). In this paper we
consider generic higher derivatives terms in the bulk
Lagrangian.  We follow the prescription of \cite{BD1} to construct an effective action
$"S_{\rm eff}"$ for transverse graviton in canonical form in presence of
generic higher derivative terms in the bulk. The effective action and original action
give same equation of motion perturbatively in the coupling of the higher derivative
terms.

Let us consider a gravity set-up with $n$ derivative action.
\be\label{gacnhd}
{\cal I }= \nt\int d^5x \sqrt{-g} \ltb R+12 + \app R^{(n)}\rtb\
\ee
where, ${\cal R}^{(n)}$ is any $n$ derivative Lagrangian.
The metric in general is given by (assuming planar symmetry),
\ben
ds^2&=&-(h_t(r)+\app\ h_t^{(n)}(r))dt^2 + {dr^2 \over h_r(r) + \app\ h_r^{(n)}(r)} \nn\\
&&+ {1\over r}
(1+ \app\
h_s^{(n)}(r))d{\vec x}^2 \ .
\een
Substituting the background metric with fluctuation (\ref{petmet}) in action
(\ref{gacnhd}) (we call it general action or original action)
for the scalar
field $\ph$ we get,
\be \label{ghdacnphi2}
{\cal I}=\nt \int {d^4 k \over (2 \pi)^4} dr \sum_{p,q=0}^{n} \cA_{p,q}(r,k)
\phi^{(p)}(r,-k) \phi^{(q)}(r,k)
\ee
 where, $\phi^{(p)}(r,k)$ denotes the $p^{th}$ derivative of the field $\ph$
with respect to $r$ and $p+q\leq n$. The coefficients $\cA_{p,q}(r,k)$ in
general depends on the coupling constant $\app$. $\cA_{p,q}$ with $p+q \ge 3$
are proportional to $\app$ and vanishes in $\app \ra 0$ limit , since the terms
$\phi^{(p)} \phi^{(q)}$ with $p+q\ge 3$ appears as an effect of higher
derivative terms in the action (\ref{gacnhd}).

Up to some total derivative
terms, the general action
(\ref{ghdacnphi}) can also be written as,
\ben \label{ghdacnphi2}
{\cal I}&=&\nt \int {d^4 k \over (2 \pi)^4} dr \sum_{p=0}^{n/2}\cA_{p}(r,k)
\phi^{(p)}(r,-k) \phi^{(p)}(r,k)\nn\\
&& \hspace{4.50cm} (\rm{for} \ n \quad \rm{even})
 \nonumber \\
&=&\nt \int {d^4 k \over (2 \pi)^4} dr \sum_{p=0}^{{n-1\over 2}}\cA_{p}(r,k)
\phi^{(p)}(r,-k) \phi^{(p)}(r,k)\nn\\
&&  \hspace{4.50cm} (\rm{for} \ n \quad \rm{odd}) \ .
\een
However this action does not have canonical form. We
write an effective action for transverse graviton in canonical form,
\ben
S_{\rm eff}&=&\nt \int {d^4 k \over (2 \pi)^4} dr \bigg[ {\cal A}_1^{\rm HD}(r,k)
 \phi'(r,k)\phi'(r,-k) \nn\\
&& \qquad + {\cal A}_0^{\rm HD}(r,k) \phi(r,k)\phi(r,-k)\bigg]
\een
with some unknown function ${\cal A}_1^{HD}$ and ${\cal A}_0^{HD}$. We fix these functions by
demanding that the equations of motion obtained from the effective action and the original action
are same perturbatively in $\app$.

The generalized canonical momentum and equation of motion
are given by,
\ben
\Pi^{\rm HD}(r,k)&=& 2 {\cal A}_1^{\rm HD}(r,k) \phi'(r,k)\nn\\
\lb \Pi^{\rm HD}(r,k)\rb'&=& 2 {\cal A}_0^{\rm HD}(r,k) \phi(r,k).
\een
Once we find the effective action for the graviton,
we follow the procedure in the previous section to
obtain the flow equation for the boundary Green's
function in generic higher derivative gravity.

The boundary Green's function is given by,
\be
G_R^{\rm HD}(k_\mu)=\lim_{r \rightarrow 0}\frac{2{\cal A}_1^{\rm HD}(r,k) \phi'(r,k)\phi(r,-k)}{\phi_0(k) \phi_0(-k)},
\ee
which can be written using the definition of canonical momentum as,
\be
G_R^{\rm HD}(k_\mu)=\lim_{r \rightarrow 0}{\Pi^{\rm HD}(r,k_\mu)\over \phi(r,k_\mu)}.
\ee
Let us define a response function of the boundary theory in higher derivative theory as,
\be
\bar \chi^{\rm HD}(k_\mu,r)= {\Pi^{\rm HD}(r,k_\mu)\over  i \omega \phi(r,k_\mu)}.
\ee
Therefore the flow equation is given by,
\be\label{hdflow}
\partial_r \bar \chi^{\rm HD}(k_{\mu},r)= i \omega \sqrt{- {g_{rr} \over g_{tt}}}
\Bigg[{\bar \chi^{\rm HD}(k_{\mu},r)^2 \over \Sigma^{\rm HD}(r,k)}- {\Upsilon^{\rm HD}(r,k) \over \omega^2}\Bigg],
\ee
where we define
\ben
\Sigma^{\rm HD}(r,k)&=& - 2 {\cal A}_1^{\rm HD}(r,k_\mu)\sqrt{-{g_{rr} \over g_{tt}}}\\
\Upsilon^{\rm HD}(r,k)&=& 2 {\cal A}_0^{\rm HD}(r,k_\mu)\sqrt{-{g_{tt} \over g_{rr}}}.
\een
This is the flow equation for two point correlation function of
energy-momentum tensor in presence of generic higher derivative
term in the bulk action. Therefore integrating this equation from
horizon to asymptotic boundary one can find the higher derivative
correction to the transport coefficients at any order in frequency/momentum.

Like two derivative case here also we need to provide
a boundary condition to solve this equation. The response function $\bar \chi^{\rm HD}(k_{\mu},r)$
should be well-defined at horizon. This implies,
\be
\bar \chi^{\rm HD}({k_{\mu}},r)\Bigg|_{r=r_h}= \sqrt {{\Sigma^{\rm HD}(r) \Upsilon^{\rm HD}(r)} \over \omega^2}\Bigg|_{r=r_h}
\ee
here the horizon is located at $r=r_h$.

One important point to mention here is that unlike two derivative
gravity where $\bar \chi({k_{\mu}},r_h)$ was independent of $k_{\mu}$,
$\bar \chi^{\rm HD}(k_{\mu},r_h)$ can in general depend on $k_{\mu}$.
We will see this explicitly in the next section. Therefore the full momentum response
at the horizon may not be able to correspond only to the zero momentum
limit of boundary response in higher derivative theory.

Like two derivative case, the response function
in higher-derivative gravity theory also contains UV divergences. We need
to add proper counter term following the holographic renormalization
procedure to cancel these divergences. A little more thinking also says
that in presence of any higher-derivative term in the action the structure
of the counterterm remains same as (\ref{CT}). Only the overall normalization
 constant depends on higher-derivative coupling. Thus, similar
to the leading gravity, the counterterm in higher derivative
gravity also cancels out the divergence and does not add any finite
contribution to the boundary response function. One can study the flow
equation of the un-renormalized response function and read off the
transport coefficients from its finite piece.

%%%%%%%%%%%%%%%%%%%%%%%%%%%%%%%%%%%%%%%%%%%%%%%%%%%%%%%%%%%%%%%%%%%%%%%
%%%%%%%%%%%%%%%%%%%%%%%%%%%%%%%%%%%%%%%%%%%%%%%%%%%%%%%%%%%%%%%%%%%%%%%

\sectiono{Examples: String theory corrections to flow equation} \label{hdcsec}

String theory predicts next to leading order corrections to Einstein-Hilbert
action. These corrections are relevant at a distance comparable with typical
length scale of the theory $l_s=\sqrt{\app}$.
The short distance corrections to this action
is described by
supplementing this action by higher curvature terms. However,
here we treat the stringy effects perturbatively $i.e$ the
coupling of higher derivative terms to be small.

From the point of view of AdS/CFT the small $\app$ correction in supergravity
corresponds
to ${1\over \lambda}$ correction in strongly coupled gauge theory
in planar limit where $\lambda$ is  the 't Hooft coupling.
% $\lambda$ in planer limit.
The precise dictionary
between string length and 't Hooft coupling is,
\be
\app^2= {L^4\over 4 \pi \lambda}\ ,  \ \ \ \ \ \ \ L \ \ AdS \ \rm{radius}.
\ee

In this section, we will consider two examples of higher-derivative
terms coming from string theory and study their effects on flow equations and second order
transport coefficients.

\subsection{$Weyl^4$ term} \label{r4sec}

 We  consider the well known $Weyl^4$
term. This term appears in type II string theory. Adding this term in the
bulk action corresponds to ${1\over \lambda^{3/2}}$ correction in dual
large $N$ theory. The string theory correction to second order
transport coefficients have already been computed in
\cite{buchel-paulos1} using usual $Kubo$ formula. Here, we
show that one can obtain the correct result by studying the $first \ order$
flow equation in higher derivative gravity without solving any second order
differential equation for graviton.

The five dimensional bulk
action is given by,
\be \label{w4acn}
S=\nt \int d^5x \sqrt{-g} \lb R + 12 + \gamma W^{(4)} \rb
\ee
where,
the coupling constant $\gamma$ is given by,
\be
\gamma = \frac{1}{8}\zeta(3)\app^3
\ee
and
\ben
 W^{(4)}&=&C^{hmnk}C_{pmnq}C_h^{\hhp rsp}C^q_{\hhp rsk}\nn\\
&& \quad+{1\over
2}C^{hkmn}C_{pqmn}C_h^{\hhp rsp}C^q_{\hhp rsk}
\een
and the Weyl tensors $C_{abcd}$ are given by,
\ben
C_{abcd}&=&R_{abcd} + {1 \over 3}
(g_{ad}R_{cb}+g_{bc}R_{ad}-g_{ac}R_{db}-g_{bd}R_{ca})\nn\\
&& \qquad \ +{1 \over
  12}(g_{ac}g_{bd}-g_{ad}g_{cb})R\ .
\een

The background metric is given by \cite{gkt,dg} \footnote{In this particular
example we keep the extremality parameter $r_0$ explicitly. It would help
us to write the relation between different transport coefficients. One can
set $r_0$ to be one by time re-scaling what we have done in the next example.},
\ben\label{w4met}
ds^2 &=& -\lb \frac{\left(1-r^2\right) r_0^2}{r}-15 r \left(3 r^6-8
   r^4+5\right) \gamma  r_0^2\rb  dt^2 \nonumber \\
&&+ \lb \frac{1}{4 r^2-4 r^4}+\frac{\left(-285 r^4+75 r^2+75\right) \gamma }{4-4
   r^2} \rb  dr^2 \nn\\
&& + {1\over r} d{\vec x}^2\ .
\een

The temperature of this black hole is given by,
\be
T= {r_0 \over \pi} \lb 1 + 15 \gamma \rb \ .
\ee
and the horizon is located at $r=1$.

The effective action for transverse graviton has been computed in \cite{BD1}.
Here we will present the result only. The effective action is given by,
\ben
S_{\rm eff} &=& \nt \int {d^4 k \over (2 \pi)^4} dr \bigg[ {\cal A}_1^{W^4}(r,k)
 \phi'(r,k)\phi'(r,-k) \nn\\
&& \qquad \ \ \ + {\cal A}_0^{W^4}(r,k) \phi(r,k)\phi(r,-k)\bigg]
\een
where, ${\cal A}_1^{W^4}$ and ${\cal A}_1^{W^4}$ are given in appendix \ref{weylapp}.

Therefore the flow equation is given by,
\be\label{W4flow}
\partial_r \bar \chi^{W^4}(k_{\mu},r)= i \omega \sqrt{- {g_{rr} \over g_{tt}}}
\Bigg[{\bar \chi^{W^4}(k_{\mu},r)^2 \over \Sigma^{W^4}(r,k)}- {\Upsilon^{W^4}(r,k) \over \omega^2}\Bigg],
\ee
where,
\ben
\Sigma^{W^4}(r,k)&=& - 2 {\cal A}_1^{W^4}(r,k_\mu)\sqrt{-{g_{rr} \over g_{tt}}}\\
\Upsilon^{W^4}(r,k)&=& 2 {\cal A}_0^{W^4}(r,k_\mu)\sqrt{-{g_{tt} \over g_{rr}}}.
\een
The explicit expressions for $\Sigma$ and $\Upsilon$ can be obtained by using ${\cal A}_1^{W^4}$ and ${\cal A}_1^{W^4}$.

From the regularity of $\bar \chi^{W^4}(k_{\mu},r)$ at horizon we get,
\ben\label{bcW4}
\bar \chi^{W^4}(k_{\mu},1) &=& {\sqrt {\Sigma^{W^4}(1) \Upsilon^{W^4}(1) \over \omega^2}}\nn\\
&=& {r_0^3\over 16 \pi G_5} +  {\gamma  r_0\over 4 \pi G_5} \left(45 r_0^2+11 q^2\right)\ .\nn\\
\een
Here, we see that unlike the two-derivative gravity, the horizon value
of the response function depends on spatial momenta $q$ (see Fig. \ref{fig2}).
With this boundary condition we solve the flow equation up to order
$\omega^2$ and $q^2$ (ignoring ${\cal O}(\omega q^2)$ term). Here we write the final
result\footnote{$k=\{\omega,0,0,q\}$ and we ignore the UV divergence piece.}.
\ben\label{chiw4}
i \omega \ \bar \chi^{W^4}(k_{\mu},0)&=& -i (1+180 \gamma ) {r_0^3 \over 16 \pi G_5} \omega\nn\\
&& + \bigg[ \frac{1}{2} (1-\log
   (2)) \nn\\
&& \ \ \ \ +\frac{5}{4} \gamma  (199-66 \log (2))\bigg] {r_0^2\over 16 \pi G_5} \omega^2\nn\\
&& - {1 \over 2}(1+20 \gamma)
{r_0^2\over 16 \pi G_5} q^2\nn\\
&& + \higho \ .
\een
Comparing this result with (\ref{Gdef}) we get
\ben
{\eta \over \pi^3 T^3} &=& 1+ 135 \gamma+ {\cal O}(\gamma^2) \nn\\
\kappa &=& {\eta \over \pi T}\lb 1- 145 \gamma\rb + {\cal O}(\gamma^2)\nn \\
\tau_{\pi} T &=& \frac{2-\log (2)}{2 \pi } + \frac{375 \gamma }{4 \pi }+ {\cal O}(\gamma^2)\ .
\een
These results are in agreement with \cite{buchel-paulos1}. It provides a
non-trivial check to this approach of obtaining higher order transport coefficients
from the flow equation (\ref{flow}).

%%%%%%%%%%%%%%%%%%%%%%%%%%%%%%%%%%%%%%%%%%%%%%%%%%%%%%%%%%%%%%%%%%%%%%%%%%%%%%%%%
%%%%%%%%%%%%%%%%%%%%%%%%%%%%%%%%%%%%%%%%%%%%%%%%%%%%%%%%%%%%%%%%%%%%%%%%%%%%%%%%%
\subsection{Four derivative term}

In this section, we will concentrate on the generic four derivative
corrections to Einstein-Hilbert action. These terms arise in the effective
action for the heterotic string theory. In fact, the complete super-symmetrized
$R^2$ correction to effective Heterotic string theory is known and one way to
obtain it is the super-symmetrization of the Lorentz Chern-Simons terms \cite{Berg,panda}.
This terms also arises in the context of Type IIB string theory \cite{Kats,Blau},
where the theory is on $AdS_5 \times X^5$, the compact space $X^5$ being $S^5/Z_2$.
The dual theory is ${\cal N}=2 Sp( N)$gauge theory with 4 fundamental and 1 antisymmetric
traceless hyper-multiplets. This super-conformal theories arises in the context
of $N$D3-branes sitting inside 8 D7-branes coincident on an orientifold 7-plane.
In this case, generic four derivative $R^2$ correction comes form the DBI action of the branes.

Here we compute the generic four derivative correction to the second order transport
coefficients, the relaxation time $\tau_{\pi}$ and $\kappa$. We can choose
the coefficients of the higher derivative terms to be the four-dimensional Euler density
and get pure Gauss-Bonnet correction to these coefficients.

The action
\ben \label{6dacnreqg}
{\cal I} &=& \nt \int d^5x \sqrt{-g} \bigg [ R +12 +
 \app  \bigg (  \beta_1 R^2 \nn\\
&& \ \ \ \ + \beta_2
R_{\mu\nu\rho\sigma} R^{\mu\nu\rho\sigma} + \beta_3 R_{\mu\nu}R^{\mu\nu}  \bigg ) \bigg].
\een

In particular for Gauss-Bonnet correction,
$\beta_1=1, \beta_2=1, \beta_3=-4.$ One can get rid
of the $Ricci^2$ and $Scalar^2$ terms by a field
redefinition and therefore all physical quantities
should depend on the coefficient $\beta_2$ only. Here,
we prefer to work with the generic case as it would be
easier for us the get the results for pure Gauss-Bonnet
combination at every step.

The background solution is given by \cite{BD2},
\be \label{6derimet}
ds^2 =  f(r) dt^2 + {g(r)\over 4 r^3} dr^2 + {1\over r} d\vec{x}^2
\ee
where $f(r)$ and $g(r)$ are given by,
\ben
f(r)&=&r - \frac{1}{r} -
 2 r\big (r^2 - 1 \big)\beta_2 \app
\een
and
\ben
g(r) &=&  \frac {r} {1- r^2} + \frac {2 r \big (
      10 \beta_1 + (1- 3 r^2) \beta_2 +
      2 \beta_3 \big) \app} { 3(r^2 - 1)}
      \ . \nn \\
\een
This is the background metric corrected up to order $\app$.
We have fixed the integration constant such that the
boundary metric is Minkowskian and the
horizon is located at $r=1$. The temperature of the
black brane is given by,
\ben\label{temp}
T&=& {1 \over \pi} +{ 10 \beta_1-5 \beta_2+ 2 \beta_3 \over 3 \pi} \app.
\een

Similar to the $Weyl^4$ case, we can write the following effective action for this model,
\ben
S_{\rm eff} &=& \nt \int {d^4 k \over (2 \pi)^4} dr \bigg[ {\cal A}_1^{GB}(r,k)
 \phi'(r,k)\phi'(r,-k) \nn\\
&& \qquad \ \ \ + {\cal A}_0^{GB}(r,k) \phi(r,k)\phi(r,-k)\bigg]
\een
where, ${\cal A}_1^{GB}$ and ${\cal A}_0^{GB}$ are given in appendix \ref{gbapp}.
 Now, it is straightforward to write the corresponding flow equation (\ref{flow}) in this case,
\be\label{GBflow}
\partial_r \bar \chi^{\rm GB}(k_{\mu},r)= i \omega \sqrt{- {g_{rr} \over g_{tt}}}
\Bigg[{\bar \chi^{\rm GB}(k_{\mu},r)^2 \over \Sigma^{\rm GB}(r,k)}- {\Upsilon^{\rm GB}(r,k) \over \omega^2}\Bigg],
\ee
where we define
\ben
\Sigma^{\rm GB}(r,k)&=& - 2 {\cal A}_1^{\rm GB}(r,k_\mu)\sqrt{-{g_{rr} \over g_{tt}}}\\
\Upsilon^{\rm GB}(r,k)&=& 2 {\cal A}_0^{\rm GB}(r,k_\mu)\sqrt{-{g_{tt} \over g_{rr}}}.
\een
%These coefficients can be obtained once ${\cal A}_1^{GB}$ and ${\cal A}_0^{GB}$ are
%known.
Now, the boundary condition (\ref{bocon}) takes the following form,
\be\label{GBbc}
\bar \chi^{GB}(k_{\mu},1)=\nt \ltb 1+\left(\left(q^2-8\right) \beta _3-40 \beta _1\right) \alpha '\rtb\ .
\ee
As mentioned earlier, we see that even in this case, the boundary condition
depends on spatial momenta $q$ through the coefficient $\beta_3$.
With this boundary condition, one can solve the flow equation (\ref{GBflow}) and the solution is given by,
\ben\label{chigb}
i \omega \ \bar \chi^{GB}(k_{\mu},0)&=& \nt\bigg[-i(1-  \left(40 \beta _1+8 \beta _3\right) \alpha ')\omega\nn\\
&&+\frac{\omega^2}{2}
   \bigg[(1-\log2)
 +\frac{\app}{6}(130 \beta _1 (\log2-1)\nn\\
&&-\beta _2 (5
   \log2-2)
   +26 \beta _3 (\log2-1))\bigg] \nn\\
&&-\frac{q^2}{2}\bigg[1 - \frac{1}{3} (130 \beta _1+25 \beta _2+26 \beta _3)
   \alpha'\bigg]\bigg]\nn\\
   && + \higho \ .
\een
From this expression we get the following transport coefficients,
\be
\eta= \nt \lb 1-  8\left(5 \beta _1+ \beta _3\right) \alpha '\rb + {\cal O}(\app^2).
\ee
This matches with results in \cite{Kats,sd,myers0}. The higher order coefficients are,
\ben\label{rgbg}
\kappa &=& {\eta \over \pi T}\lb 1- 10 \beta_2 \app\rb + {\cal O}(\app^2)\nn \\
\tau_{\pi} T &=& {2-\ln2 \over 2 \pi}- {11 \beta_2 \over 2\pi}\app + {\cal O}(\app^2) .
\een

As we can see, the physical quantities $\eta/s, \kappa,\tau_{\pi} T$ only
depend on the coefficient $\beta_2$. In particular to Gauss-Bonnet combination, the corrections are,
\ben\label{GBTC}
\kappa &=& {\eta \over \pi T}\lb 1- 10 \app\rb + {\cal O}(\app^2)\nn \\
\tau_{\pi} T &=& {2-\ln2 \over 2 \pi}- {11 \over 2 \pi}\app+ {\cal O}(\app^2).
\een
(\ref{rgbg}) and (\ref{GBTC}) are new results of this paper.

\subsubsection{Exact result for Gauss-Bonnet black hole}

As we have done the above computation perturbatively, the above
 expressions are valid only at order $\alpha'$. But, one can
  consider the Gauss-Bonnet term exactly in coupling. For pure
  Gauss-Bonnet combination the equations of motion remain second order
  differential equation and hence it is easy to solve exactly to find
  the background space-time.
  We solve the flow equation in this background exactly in coupling constant,
  and find the exact expressions
  for relaxation time $\tau_{\pi}$ and $\kappa$. In this section we briefly
  outline the result.

  The action and the solution is given by,
\ben
{\cal I}_{GB}&=&\nt\int d^5x \sqrt{-g}\bigg [ R +12 \nn\\
&& \ \ +
{\lambda_{gb}\over 2}  \bigg (R^2 +
R_{\mu\nu\rho\sigma} R^{\mu\nu\rho\sigma}  -4 R_{\mu\nu}R^{\mu\nu}  \bigg ) \bigg]\nn\\
ds^2&=& r^2\lb -{f(r)\over f_{\infty}} dt^2 + d\vec x^2\rb + {dr^2 \over r^2 f(r)}
\een
where,
\be
f_{\pm} = {1\over 2 \lambda_{gb}}\ltb 1 +\pm  \sqrt{1-4\lambda_{gb}\lb 1- {r_0^4\over r^4}\rb}\rtb
\ee
and
\be
f_{\infty}=\lim_{r\ra \infty} f(r) = {1-\sqrt{1-4 \lambda_{gb}}\over 2 \lambda_{gb}}\ .
\ee

In this coordinate the boundary metric is $\eta$. We also consider only the $'-'$ branch of $f_{\pm}$ which
corresponds to a non-singular black hole solution with non-degenerate horizon.

The black hole temperature is given by,
\be
T={r_0\over \pi f_{\infty}}\ .
\ee

With exact GB, the effective action for fluctuation has a canonical form.
Therefore we derive the flow equation for the response function (as we did
in section \ref{model}) and solving this equation we get,
\ben\label{exact}
\eta &=& \nt (1- 4 \lambda_{gb})\nn\\
\kappa &=& \frac{2 \lambda _{gb} \left(8 \lambda
   _{gb}-1\right)}{\left(1-\sqrt{1-4 \lambda
   _{gb}}\right) \left(4 \lambda _{gb}-1\right)}\nn \\
\tau_{\pi}T &=& {1\over 4\pi (-1 + 4 \lambda_{gb})}\bigg[-8 \lambda_{gb}^2+12 \sqrt{1-4 \lambda_{gb}}
   \lambda_{gb}\nn\\
   && +10 \lambda_{gb}-2 \sqrt{1-4
   \lambda_{gb}}
    -4 \log (2) \lambda _{gb}\nn\\
    && +\left(1-4
   \lambda_{gb}\right) \log \left(-4 \lambda_{gb}+\sqrt{1-4 \lambda_{gb}}+1\right)\nn\\
   &&+\left(4
   \lambda_{gb}-1\right) \log \left(1-4 \lambda_{gb}\right)-2+\log (2)\bigg].\nn\\
\een
One can easily check that up to first order in $\lambda_{gb}$, the results in
(\ref{exact}) reduces to the one in (\ref{GBTC}). In \cite{myers1} the authors
obtained the relation between second order transport coefficients and $\lambda_{gb}$
numerically, however we are able to present the result exactly.

As we have mentioned in introduction that the flow equation is a
first order non-linear differential equation but one can reduce
this equation to a second order linear differential equation. This
second order differential equation is related to the equation of
motion for transverse graviton (gauge invariant excitations). Therefore
we can use this equation to study causality violation in Gauss-Bonnet
gravity. In \cite{myers1,myers2} it was found that to preserve causality of a conformal
fluid there exists a bound on second order transport coefficients,
\be
\tau_{\pi}T-2 {\eta\over s}\geq 0\ .
\ee
In Fig \ref{fig3} we plot $\tau_{\pi}T-2 {\eta\over s}$ for our result
and find the following bound on $\lambda_{gb}$
which is in agreement with \cite{myers1}.
\be
-0.711\leq \lambda_{gb}\leq 0.113\ .
\ee

\lfig{Bound on $\lambda_{gb}$.}
{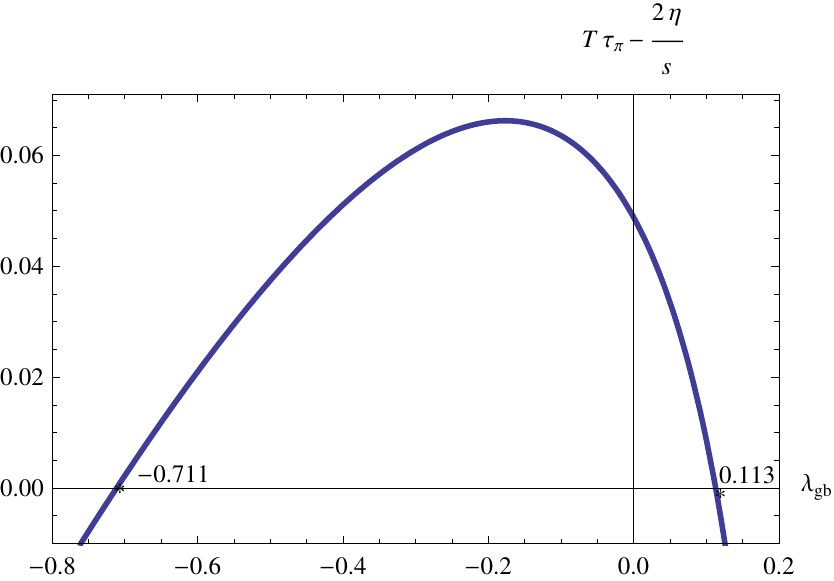}{6.5cm}{fig3}

%%%%%%%%%%%%%%%%%%%%%%%%%%%%%%%%%%%%%%%%%%%%%%%%%%%%%%%%%%%%%%%%%%%%%%%%%%
%%%%%%%%%%%%%%%%%%%%%%%%%%%%%%%%%%%%%%%%%%%%%%%%%%%%%%%%%%%%%%%%%%%%%%%%%%%%%
%%%%%%%%%%%%%%%%%%%%%%%%%%%%%%%%%%%%%%%%%%%%%%%%%%%%%%%%%%%%%%%%%%%%%%%%%%%%%
%%%%%%%%%%%%%%%%%%%%%%%%%%%%%%%%%%%%%%%%%%%%%%%%%%%%%%%%%%%%%%%%%%%%%%%%%%%%%

\sectiono{Flow equation for charged black holes}\label{cbh}

Electrically charged black holes in five dimensions have drawn a lot of interest in the
context of AdS/CFT. The electric charge of these black holes are mapped to the global
R-charge of the dual field theory. Because of the presence of the electric charges, the
thermodynamics and the phase structure of these black holes are rather complicated and also
interesting at the same time. There have been a lot of study of thermodynamics and phase
transitions of these charged black hole with different horizon topologies (see \cite{Cvetic} and
references therein).

The goal of the present section is to apply the AdS/CFT correspondence to understand
how non-vanishing chemical potentials effect the hydrodynamic behavior of strongly coupled gauge
theories.
We study the second order hydrodynamics in two cases:
(a) Generic R-charge black holes and (b) Charge black holes in
higher-derivative gravity.

\subsection{R-charged black holes}

We consider a conformal field theory with conserved charge (density) in addition to energy and
momentum. This is especially an interesting extension of the hydrodynamics of the uncharged fluids.

The second order hydrodynamics of charged fluid has been studied in \cite{chargehydro,haack}.
They consider Reissner-Nordstrom black hole in five dimensions and found the effect of chemical
potential on second order transport coefficients in some limits of chemical potential. However, we consider
generic $R$-charged black holes with three (unequal) charges (chemical potentials) and find
the exact expressions for second order transport coefficients in presence of three chemical potentials. As we
have mentioned in the introduction that solving the flow equation (of retarded Green's function
of energy momentum tensor) we can only find two second
order transport coefficients whereas in \cite{chargehydro, haack} all other second order transport
coefficients have been reported.

%The goal of this present section is to investigate, using the flow equation, how a non-vanishing chemical
%potential effects the second order transport coefficients.

We consider R-charged black holes in five dimensions. A consistent truncation of ${\cal N}=8$, $D = 5$ gauged supergravity
with $SO(6)$ Yang-Mills gauge group, which can be obtained by $S^5$ reduction of type $IIB$
supergravity, gives rise to ${\cal N} = 2$, $D = 5$ gauge supergravity with $U(1)^3$ gauge group. The
same theory can also be obtained by compactifying eleven dimensional supergravity, low
energy theory of $M$ theory, on a Calabi-Yau three folds. The bosonic part of the action of
${\cal N} = 2$, $D = 5$ gauged supergravity is given by \cite{Cvetic}. We follow the notation of \cite{son5}.
\ben
{\cal I}_{\rm sugra} &=& \nt \int d^5x \sqrt{-g} \bigg[R +V(X) \nn\\
&&- {1 \over 2}G_{IJ}(X) F^I_{\mu\nu}F^{\mu\nu J}-G_{IJ}(X)
\partial_{\mu} X_I \partial^{\mu}X_J \bigg]  \nonumber \\
&& \quad +{{\zeta\over 3} \over 16\pi G_5} \int  d^5x\ \epsilon^{\mu\nu\rho\sigma\gamma}A_{\mu} F_{\nu\rho} F_{\sigma\gamma}
\een
where, $X^I$'s are three real scalar fields, subject to the constraint
$X^1 X^2 X^3 =1$. $F^I$'s, which are field strengths of three Abelian gauge
fields (I,J=1,2,3), and the scalar potential $V(X)$ is given by,
\ben
F_{\mu\nu}^I &=& 2 \partial_{[\mu}A^I_{\nu]}\nn\\
G_{IJ}&=&{1\over 2} \rm{diag} \ltb (X^1)^{-2},(X^2)^{-2},(X^3)^{-2}\rtb\nn\\
V(X)&=&2\sum_I{1\over X_I}
\een

The three-charge non-extremal STU solution is specified by
the following background values of the metric
\ben
ds^2 &=& - {\cal H}^{-2/3}\, f_k \, dt^2
+ {\cal H}^{1/3} \left( f_k^{-1} dr^2 + r^2 d \Omega_{3,k}^2\right)\,,\nn
\\
\label{metric_k}
\een
%
%\begin{equation}
%ds^2 = - \left( H_1 H_2 H_3\right)^{-2/3}\, f_k \, dt^2
%+ \left( H_1 H_2 H_3\right)^{1/3} \left( f_k^{-1} dr^2
% + r^2 d \Omega_{3,k}^2\right)\,,
%\end{equation}
\ben
f_k = k - {m_k\over r^2} + {r^2}{\cal H} \,, \qquad
 H_i = 1 + {q_i\over r^2}\,, \nn\\
  {\cal H} = H_1 H_2 H_3 \,, \hspace{2cm}
\een
 as well as the scalar and the gauge fields
\begin{equation}
X^i = {{\cal H}^{1/3}\over H_i} \,,  \qquad
A^i_t = \sqrt{ {k q_i + m_k\over q_i}} \left( 1 - H_i^{-1}\right)\,.
\end{equation}
The parameter $k$ determines the spatial curvature of  $d \Omega_{3,k}^2$:
$k=1$ corresponds to the metric on the three-sphere of unit radius,
$k=0$ - to the metric on $R^3$.
%It was shown in \cite{Cvetic:1999ne} that the $k=0$ solution arises as the
% Kaluza-Klein reduction on $S^5$ of the ten-dimensional metric
%describing spinning near-extremal three-branes. The three R-charges
% $q_i$ are related to the three independent angular momenta in ten
% dimensions.
Hydrodynamic approximation is valid only in the
case of a translational-invariant horizon, in our case we set
$k=0$
and
$$
d \Omega_{3,0}^2 \rightarrow  \left( dx^2 + dy^2 + dz^2\right)\,.
$$
Replacing the radial coordinate $r \ra r_0/\sqrt{r}$,
where $r_0$ is the largest root of the equation $f(r)=0$,
the background solution in this new coordinate is given by,
\ben
ds^2_5 &=& - {\cal H}^{-2/3}{(\pi T_0)^2 \over r}\,f \, dt^2 + {\cal H}^{1/3}{1 \over 4 f r^2} dr^2\nn\\
&+&  {\cal H}^{1/3}{(\pi T_0 )^2 \over r}\, \left( dx^2 + dy^2 + dz^2\right)
\,,
\label{metric_u_3}
\een
\ben
f(r) &=& {\cal H} (r) - r^2 \prod\limits_{i=1}^3 (1+\kappa_i)\,,
\;\;\;\;\; H_i = 1 + \kappa_i r \,, \nn\\
&& \hspace{2cm} \kappa_i \equiv {q_i\over r_0^2}\, .
\label{identif}
\een
where $\kappa_i's$ are chemical potentials and
\be
 T_0 = r_0/\pi\,.
\ee
The scalar fields and the
gauge fields are given by
%\footnote{We change the normalization of
%the gauge fields by a factor of $\sqrt{2}/L$.
%This normalization is used in the rest of the paper.}
\begin{equation}
X^i = {{\cal H}^{1/3}\over H_i(u)} \,, \qquad
A^i_t = {\tilde{\kappa}_i \sqrt{2} u\over L H_i(u)}
\label{scal_gauge_u_3}
\end{equation}
where,
\be
\tilde{\kappa}_i = {\sqrt{q_i}}
 \prod\limits_{i=1}^3 (1+\kappa_i)^{1/2}
\,.
\ee
The Hawking temperature of the background
(\ref{metric_u_3}) is given by
\begin{equation}
T_H =
{2 + \kappa_1 + \kappa_2 + \kappa_3 - \kappa_1 \kappa_2  \kappa_3\over
2\sqrt{(1+\kappa_1)(1+\kappa_2) (1+\kappa_3)}}\, T_0\,.
\end{equation}

We perturb the $xy$ component of background metric and the action
for transverse graviton is given by,
\ben
S_{\rm eff} &=& \nt \int {d^4 k \over (2 \pi)^4} dr \bigg[ {\cal A}_1^{Q}(r,k)
 \phi'(r,k)\phi'(r,-k) \nn\\
&& \qquad \ \ \ + {\cal A}_0^{Q}(r,k) \phi(r,k)\phi(r,-k)\bigg]
\een
where,
\be
{\cal A}_1^{\rm Q}= -{r_0^4 f(r) \over r}
\ee
and
\ben
{\cal A}_0^{\rm Q}   &=& {r_0^2\over 4r^2} \lb {H_1 H_2 H_3\over f(r)} -q^2\rb\ .
\een

Therefore the flow equation is given by,
\be\label{Rflow}
\partial_r \bar \chi^{\rm Q}(k_{\mu},r)= i \omega \sqrt{- {g_{rr} \over g_{tt}}}
\Bigg[{\bar \chi^{\rm Q}(k_{\mu},r)^2 \over \Sigma^{\rm Q}(r,k)}- {\Upsilon^{\rm Q}(r,k) \over \omega^2}\Bigg].
\ee

Solving this equation perturbatively in $\omega$ and $q$ we get
\ben
\bar \chi^{\rm Q}(k_{\mu},r)&=&-\frac{r_0^3\prod_i(1+\kappa_i)^{1/2}}{16\pi G_5}\nn\\
&& +{i r_0^2\over 2 \omega}\frac{(q^2-\omega^2)}{16\pi G_5}\lb1-{1\over r}\rb\nn\\
&&+{i \omega r_0^2 \prod_i(1+\kappa_i) \over 16\pi G_5\sqrt{4 P_{\kappa} +(1+S_{\kappa})^2}}\nn\\
&\bigg(&{1\over 2} \ln \ltb \frac{1+ S_{\kappa} -2 P_{\kappa}-\sqrt{4 P_{\kappa} +(1+S_{\kappa})^2}}{1+ S_{\kappa} -2 r S_{\kappa}-\sqrt{4 P_{\kappa} +(1+S_{\kappa})^2}}\rtb\nn\\
&+&{1\over 2} \ln \ltb \frac{1+ S_{\kappa} -2 r P_{\kappa}+\sqrt{4 P_{\kappa} +(1+S_{\kappa})^2}}{1+ S_{\kappa} -2 S_{\kappa}+\sqrt{4 P_{\kappa} +(1+S_{\kappa})^2}}\rtb\bigg)\nn\\
&& + {\cal O}(q\omega^2, \omega q^2, q^3, \omega^3)
\een
where,
\ben
S_{\kappa}&=&\sum_i \kappa_i\nn\\
P_{\kappa}&=&\prod_i\kappa_i.
\een

Computing the response function at the boundary (throwing away the divergent piece)
 we get the following transport coefficients,
\ben
\eta &=& \frac{r_0^3}{16\pi G_5}\prod_i(1+\kappa_i)^{1/2}
\een
and
\ben
\kappa &=& {\eta\over \pi T} {1 +S_{\kappa}/2-P_{\kappa}/2 \over  \prod_i (1+\kappa_i)}\nn\\
\tau_{\pi}T&=& {2 +S_{\kappa}-P_{\kappa}\over 4 \pi \prod_i(1+\kappa_i)} \bigg[ 2- {\prod_i(1+\kappa_i)\over \sqrt{4 P_{\kappa}+(1+S_{\kappa})^2}}\nn\\
&& \ln\lb {3+S_{\kappa} + \sqrt{4 P_{\kappa}+(1+S_{\kappa})^2}}\over 3+S_{\kappa} - \sqrt{4 P_{\kappa}+(1+S_{\kappa})^2}\rb\bigg].
\een
These are the new results in this paper. It is easy to check that for $\kappa_i \ra 0$ limit
we recover the results in section \ref{model}.

To complete the discussion on the second order transport coefficients for
$R$-charged black holes one should find the flow of Green's functions for two
point correlation functions of $R$-currents. As we will mention in section \ref{rcflow}
that in presence of finite charges (or chemical potentials) it is very hard to
solve the $Riccati$ equation even perturbatively in $\omega$ and $q$. We find
it very difficult to get any analytic solution for $R$-current Green's function.
However we consider a simple model in section \ref{rcflow} and study the flow of
$R$-current Green's function numerically.

%%%%%%%%%%%%%%%%%%%%%%%%%%%%%%%%%%%%%%%%%%%%%%%%%%%%%%%%%%%%%%%%%%%%%%%%%%%%%%%%%%%%%%%%%%%%%%%%%
%%%%%%%%%%%%%%%%%%%%%%%%%%%%%%%%%%%%%%%%%%%%%%%%%%%%%%%%%%%%%%%%%%%%%%%%%%%%%%%%%%%%%%%%%%%%%%%%%%%

\subsection{Charged black holes in higher derivative gravity}

In this section, we will study five-dimensional gravity in presence of a negative
cosmological constant and coupled to $U(1)$ gauge field. The model has been studied in \cite{mps,Maeda,sl}, the action is given as,
\ben\label{hdcbacn}
S&=&\frac{1}{16 \pi G_5} \int d^5x \sqrt{-g}\bigg [R +12 -\frac{1}{4}F^2 \nn\\
&& \,\,\,\,\,\,\,\,\,\,\,\,\ +
\frac{\zeta}{3} \epsilon^{abcde}A_{a}F_{bc}F_{de} + \alpha' \bigg( c_1 R_{abcd}R^{abcd} \nn \\
&& \,\,\,\,\,\,\,\,\,\,\,\,\ + c_2 R_{abcd}F^{ab}F^{cd} +c_3(F^2)^2+c_4 F^4\nn\\
&& \,\,\,\,\,\,\,\,\,\,\,\,\
+ c_5 \epsilon^{abcde}A_{a}R_{bcfg}R_{de}^{fg}\bigg) \bigg].
\een
Here, $F^2= F_{ab}F^{ab}, F^4=F_{ab}F^{bc}F_{cd}F^{da}$, and the AdS radius is set to unity.
The action includes the Chern-Simon term and also a generic set of four derivative terms.
All the four derivative terms will be treated perturbatively in our computation and here
$\alpha' << 1$ is the perturbation parameter. In \cite{mps}, it was shown that, within
perturbative approach, after using field-redefinition, this is the most generic four
derivative action that one can write down. In this section, we will closely follow their
work. The background metric and the gauge field in presence of these higher derivative
terms have the following form,
\ben\label{chargehdsol}
ds^2&=& - r^2 f(r) dt^2 + \frac{1}{r^2 g(r)}dr^2 + r^2 (dx^2+ dy^2+dz^2), \nn \\
A &=& h(r) dt,
\een
where,
\ben\label{hds}
f(r)&=&f_0(r)(1+ \alpha' F(r)), \nn \\
g(r)&=& f_0(r)(1+\alpha' (F(r)+ G(r))), \nn \\
h(r)&=&h_0(r)+\alpha'H(r).
\een
Here $f_0(r),g_0(r)$ and $h_0(r)$ are the solution of the background in absence of
the higher-derivative terms in the action and they are given as,
\ben
f_0&=&g_0=\bigg(1-\frac{r_0^2}{r^2}\bigg)\bigg(1+\frac{r_0^2}{r^2}-\frac{Q^2}{r_0^2 r^4} \bigg), \nn \\
h_0&=&{\sqrt{3}Q}\bigg(\frac{1}{r_0^2}-\frac{1}{r^2}\bigg).
\een
Here, $Q$ is related to the physical charge of the system and $r_0$ is the
 position of the horizon. From (\ref{hds}), it is clear that even in presence
 of the higher-derivative terms, the horizon remains at $r_0$. The higher-derivative
 corrections to this background are given by the functions $F(r),G(r)$ and $ H(r)$.
 The form of these functions are given in \cite{mps}. We would not write those
 expressions and refer the reader to that paper.

Using the flow equation,
 we will study the higher order transport coefficient of the plasma theory dual to
 this gravity model. For this, we will write the effective action for the metric
 fluctuation in (\ref{petmet}), as we have done in previous sections,
\ben
S_{\rm eff} &=& \nt \int {d^4 k \over (2 \pi)^4} dr \bigg[ {\cal A}_1^{CB}(r,k)
 \phi'(r,k)\phi'(r,-k) \nn\\
&& \qquad \ \ \ + {\cal A}_0^{CB}(r,k) \phi(r,k)\phi(r,-k)\bigg]
\een
where, ${\cal A}_1^{CB}$ and ${\cal A}_0^{CB}$ are given in appendix \ref{hdcb}. The
corresponding the flow equation (\ref{flow}) for this case with the coefficients
${\cal A}_1^{CB}$ and ${\cal A}_0^{CB}$ is,
\be\label{CBflow}
\partial_r \bar \chi^{\rm CB}(k_{\mu},r)= i \omega \sqrt{- {g_{rr} \over g_{tt}}}
\Bigg[{\bar \chi^{\rm CB}(k_{\mu},r)^2 \over \Sigma^{\rm CB}(r,k)}- {\Upsilon^{\rm CB}(r,k) \over \omega^2}\Bigg],
\ee
where we define
\ben
\Sigma^{\rm CB}(r,k)&=& - 2 {\cal A}_1^{\rm CB}(r,k_\mu)\sqrt{-{g_{rr} \over g_{tt}}} \nn \\
\Upsilon^{\rm CB}(r,k)&=& 2 {\cal A}_0^{\rm CB}(r,k_\mu)\sqrt{-{g_{tt} \over g_{rr}}}.
\een

We solve this flow equation to find the effect of higher derivative terms and chemical potential (or charge)
on transport coefficient. Here, we present our result for small $Q$ only, though it is possible
to find the results for any $Q$.

The boundary condition (\ref{bocon}) will take the following form\footnote{Note that we are
working in a different coordinate where $r\ra \infty$ is the boundary, therefore we choose
the positive branch of the boundary condition (\ref{bocon})},
\be
\bar \chi(k_\mu,r_0)=\frac{r_0^3}{16 \pi G_5}\bigg[1- \alpha'\frac{24 c_1 Q^2}{r_0^6}\bigg].
\ee
In this case, the horizon value of the response function is independent of
momenta. With this boundary condition, we can solve the flow equation
(\ref{CBflow}), and the solution is given by,
\ben
i \omega \bar \chi(k_\mu,\infty)&=& i \omega \frac{r_0^3}{16 \pi G_5}\bigg[1- \alpha'\frac{24 c_1 Q^2}{r_0^6}\bigg] \nn \\
&&- \omega^2\frac{r_0^2}{32 \pi G_5} \bigg [(1-\ln2) -\frac{Q^2}{2 r_0^6}(3-\ln2)\nn \\
&& + \frac{\alpha'}{6}\bigg(c_1(2-5 \ln2)
-\frac{Q^2}{2 r_0^6}(c_1(35 -58 \ln2)\nn\\
&&- 48 c_2(2-\ln2))\bigg)\bigg] \nn \\
&&+\frac{q^2 r_0^2}{32 \pi G_5}\bigg[1- \frac{\alpha'}{3}\bigg(25 c_1 +\frac{Q^2}{r_0^6}(32 c_1+24 c_2)\bigg)\bigg] \nn \\
&& + \higho + {\cal O}(Q^4)\ .
\een

It is easy to read off the transport coefficients from this expressions.
\ben
\eta&=&\frac{r_0^3}{16 \pi G_5}\bigg[1- \alpha'\frac{24 c_1 Q^2}{r_0^6}\bigg] + {\cal O}(Q^4)\nn \\
\kappa&=& \frac{\eta}{\pi T}\bigg[\lb 1-\frac{Q^2}{2 r_0^6}\rb - \alpha'\lb 10 c_1 - \frac{Q^2}{3 r_0^6}(37 c_1 -48 c_2)\rb\bigg] \nn\\
&& \qquad \qquad \hspace{5cm}+ {\cal O}(Q^4)\nn \\
\tau_{\pi} T &=&{2 - \ln 2\over 2\pi} -{Q^2(5-3 \ln 2)\over 4 \pi r_0^6}\nn\\
 && \ \ \ \ + \app \ltb -{11 c_1 \over 2 \pi} +{Q^2 \over 4 \pi r_0^6}(-16 c_2 + 5 c_1(11-4\ln 2))\rtb \nn \\
&& \qquad \qquad \hspace{5cm}+ {\cal O}(Q^4)\nn \\
\een
where, the temperature $T$ of the system is given by,
\be
T=\frac{r_0}{\pi}\bigg[\lb 1-{Q^2 \over 2 r_0^6}\rb-{\alpha' \over 3}\bigg(5 c_1 +\frac{Q^2}{2 r_0^6}(31 c_1 + 48 c_2)\bigg)\bigg]+ {\cal O}(Q^4).
\ee

We see that the first order as well as the second order transport coefficients coming
from retarded Green's function of energy momentum tensor\footnote{It would be
interesting to study the flow of retarded Green's function for boundary $R$-current.
We found it to be difficult to get any analytical solution for response function
in presence of finite chemical potential and higher derivative terms. However, it would be
nice to know the higher derivative corrections to other second order transport coefficients
appear in $R$ current\cite{chargehydro, haack}.} only depends on two coefficients
$c_1, c_2$. This feature was observed in \cite{mps} for entropy density $s$ and first-order
transport coefficients $\eta$. The coefficients $c_3, c_4$, which parameterize
couplings in the four point function
of the dual $U(1)$ current does not play any role in these hydrodynamic coefficients. They
should be important for the computation of conductivity, which
comes from the Green's function of the boundary $R$-current. Two
other coefficients $\zeta, c_5$ also do not appear in the expressions. One can find a magnetic
brane solution of the action (\ref{hdcbacn}) like \cite{kraus-dhoker}. In that case it would be interesting to
find the effect of magnetic field on transport coefficients.

%%%%%%%%%%%%%%%%%%%%%%%%%%%%%%%%%%%%%%%%%%%%%%%%%%%%%%%%%%%%%%%%%%%%%%%%%%%%%%%%%%%
%%%%%%%%%%%%%%%%%%%%%%%%%%%%%%%%%%%%%%%%%%%%%%%%%%%%%%%%%%%%%%%%%%%%%%%%%%%%%%%%%%%
%%%%%%%%%%%%%%%%%%%%%%%%%%%%%%%%%%%%%%%%%%%%%%%%%%%%%%%%%%%%%%%%%%%%%%%%%%%%%%%%%%%

\sectiono{Flow of retarded Green's function of boundary $R$ current}\label{rcflow}

Finally, in this section, we study the flow of retarded Green's function
of boundary $R$-current,
\be
G^R_{i,j}(k) =-i \int dt d^3x e^{ik\cdot x} \langle[J_i(x),J_j(0)]\rangle
\ee
where $J_{\mu}(x)$ is the $CFT$ current dual to a bulk gauge field $A_{\mu}$.

In hydrodynamic approximation one can express the current in powers of boundary derivatives.
Up to first order in derivative expansion it has the following form,
\be
J_{\nu}= - \tilde{\kappa} P^{\alpha}_{\nu} \partial_{\alpha}{\mu\over T} + \Omega l_{\nu} + {\cal O}(\partial^2)
\ee
where, $\tilde{\kappa}$ and $\Omega$ are two first order transport coefficients, $\mu$ is chemical
potential, $T$ is temperature and
\ben
P_{\mu\nu}&=&u_{\mu}u_{\nu} + \eta_{\mu\nu} \nn\\
l_{\mu}&=& \epsilon_{\mu}^{\alpha\beta\gamma}u_{\alpha}\partial_{\beta}u_{\gamma} \ .
\een
The expression of $J_{\mu}$ up to second order in derivative expansion can be found in \cite{chargehydro,haack}.
From conformal invariance of the theory it is possible to write
all possible second order transport coefficients appear in the expression of $J_{\mu}$. However,
 like energy momentum tensor, from the expression of retarded Green's function it is not possible
to compute all the transport coefficients that appear in different order of derivative expansion.

In this section we study the flow equation of retarded Green's function of boundary $R$-current.
Unfortunately we find it difficult
to solve the flow equation analytically to extract any transport coefficient. We present our calculation
how to write the flow equation for retarded Green's function of $R$-currents in presence of generic higher derivative
terms in bulk Lagrangian and some numerical results.

We start with Einstein-Maxwell action
\be
S= \nt \int d^5x \sqrt{-g}\lb R + 12 -\frac{1}{4}F^2\rb.
\ee
Solution is given by equation (\ref{chargehdsol}) with $\app=0$.
The temperature of the black hole is given by,
\be
T={r_0\over \pi}\lb 1-\frac{Q^2}{2r_0^6}\rb
\ee
and the chemical potential is given by,
\be
\mu=\frac{\sqrt{3} Q}{r_0^2}.
\ee

For technical advantage we write the metric and gauge field in
a different coordinate. We change
the radial coordinate $r\ra {r_0 \over \sqrt{r}}$. In this coordinate the metric
and gauge field is given by,
\ben
ds^2 &=& - {r_0^2 U(r)\over r} dt^2 + {dr^2 \over 4 r^2 U(r)}+ {r_0^2\over r}(d\vec{x}^2)\nn\\
A_t(r) &=& E(r)
\een
where,
\ben
U(r)&=&(1-r)(1+r-{Q^2 r^2\over r_0^6})\nn\\
E(r)&=&{\sqrt{3}Q\over r_0^2}(1-r)\ .
\een

We turn on small fluctuations for $x$ component of gauge fields.
Since the $A_t$ component of the bulk vector is non-vanishing
in this background, the perturbations $A_x$ can couple to the $tx$
component of graviton
Therefore we also need to consider small
metric fluctuations for components $g_{tx}$.
Writing them in momentum space
\ben
A_{x}(r,x)&=&\int{d^4k\over (2\pi)^4} e^{i k.x} A_{1}(r,k)\nn\\
g_t^{x}(r,x)&=&\int{d^4k\over (2\pi)^4} e^{i k.x} \Phi(r,k)\ .
\een
However, there exists a constraint relation between $A_x$ and $g_{tx}$. We
use this relation to replace $g_{tx}$ from equation of motion of $A_x$.

 The on-shell action for gauge field fluctuations
are given by,
\ben\label{Axacn}
S_{A} &=& \nt \int {d^4k\over (2 \pi)^4} \bigg[ -r_0^2 U(r) A_x'^2(r,k) \nn\\
&& \qquad +  \lb {\omega^2 \over 4 r U(r)} - {q^2\over 4 r}\rb A_x^2\bigg].
\een
The current corresponding to $A_x$ fluctuation is given by,
\ben
J_x(r,k)&=&{\delta S_{A} \over \delta A_x'(r,k)}\nn\\
&=&-2 r_0^2 U(r) A_x'(r,k).
\een
The equation of motion for $A_x(r,k)$ is given by,
\ben
(U(r) A_x'(r,k))' &=& -{1\over 4 r_0^2} \lb{\omega^2 \over r U(r)} - {q^2\over r}\rb A_x(r,k) \nn\\
 &&-E'(r) \phi'(r,k)\nn\\
 &=& -{1\over 4 r_0^2} \lb{\omega^2 \over r U(r)} - {q^2\over r}\rb A_x(r,k) \nn\\
 && + {r E'(r)^2 \over r_0^2}A_x(r,k).
\een
Using the constraint relation (coming from $rx$ component of Einstein equations),
\be
\phi'(r,k) =- {r E'(r) A_x(r,k) \over r_0^2}.
\ee
and the equations of motion one gets,
\ben
J_x'(r,k) &=& {1\over 2} \lb{\omega^2 \over r U(r)} - {q^2\over r}\rb A_x(r,k) \nn\\
&& - {2r E'(r)^2} A_x(r,k).
\een

Next we define a response function
\be
\sigma(r,k)= {J_x(r,k) \over i \omega A_x(r,k)}.
\ee
Taking the derivative with respect to $r$ and using the equation of motion we
find the flow is given by,
\ben\label{Jfloweq}
\sigma'(r,k) &=& {i \omega\over 2 r_0 U(r)} \bigg[ \sigma(r,k)^2 \nn\\
&& - \bigg({r_0\over  r} \bigg( 1 - {q^2 U(r) \over \omega^2} \bigg) - {4 r_0 r E'(r)^2 U(r) \over \omega^2} \bigg)\bigg]. \nn \\
\een
From the regularity of the response function at the horizon we find the boundary condition
is given by,
\be
\sigma(1,k)^2 = 1.
\ee

With this boundary condition one can integrate this nonlinear equation
to find finite frequency response of boundary Green's function.

In presence of generic higher derivative terms in the Lagrangian the on-shell
action for fluctuation $A_x$ may not have canonical form like (\ref{Axacn}).
In that case one has to write an effective action for $A_x$ like transverse
graviton. The effective will have the same form as (\ref{Axacn}) only the
coefficients will depend on coupling constant of higher derivative terms.

We conclude this section by presenting some numerical solutions of the
flow equation (\ref{Jfloweq}) for both non-extremal extremal black holes in
fig. \ref{fig4}.

\lfig{Flow of $\sigma$ for non-extremal and extremal black holes.}
{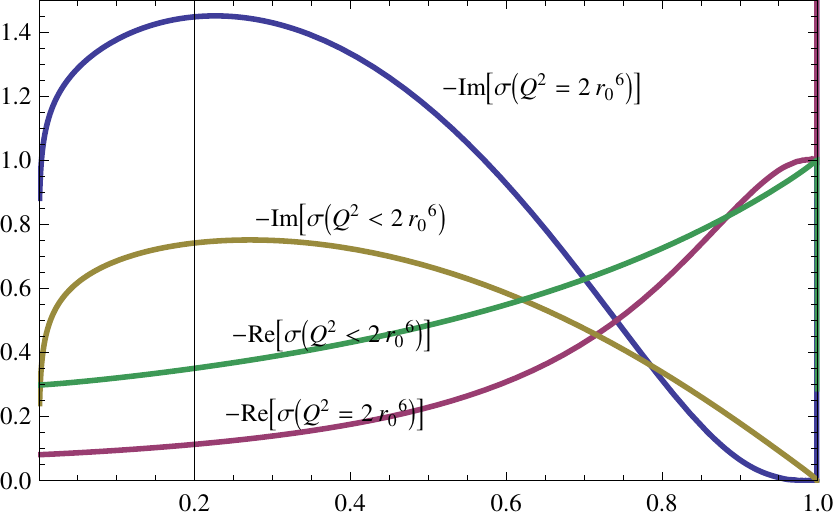}{6.5cm}{fig4}

\begin{acknowledgments}
We would like to thank Dileep Jatkar and Umut Gursoy for useful discussions.
The work of NB is part of the research program of the Foundation for
Fundamental Research on Matter (FOM), which is financially supported
by the Netherlands Organization for Scientific Research (NWO).
\end{acknowledgments}

\appendix

\section{Second order transport coefficients from Kubo formula}\label{kubo}

In this appendix we re-derive second order transport coefficients $\kappa$ and $\tau_{\pi}$
 from usual Kubo approach.
Let us now consider the action with solution given in section \ref{model},
\be\label{action}
S={1 \over 16 \pi G_5}\int d^5x\sqrt{-g}  [R + 12 ]\, .
\ee
For well defined variation of this action we need to add a Gibbons-Hawking
boundary term. Also, requiring the on-shell action
being finite at boundary, we have to add counter-terms following usual approach
of holographic renormalization. They are as follows:
\ben
S_{GH}&=&{1 \over 8 \pi G_5}\int d^4x\sqrt{-\gamma} {\cal K},\nn \\
S_{CT}&=& {1 \over 16 \pi G_5}\int d^4x\sqrt{-\gamma}[6+ {1 \over 2} {\cal R}]
\een
where $\gamma$ and ${\cal R}$ are boundary metric and Ricci scalar (constructed out of $\gamma$)
respectively.

We consider the following metric perturbation,
\be
g_{xy}=g^{(0)}_{xy}+ h_{xy}(r,x)=g^{(0)}_{xy}(1+\ep \Phi(r,x)),
\ee
where $\epsilon$ is an order counting parameter. We are interested in quadratic on-shell
action for this transverse graviton  $\Phi(r,x)$. Let us define the Fourier transform,
\be \label{phifu}
\ph = \int {d^4x \over (2 \pi)^4} e^{-i k.x} \Phi(r,x)\ ,
\ee
and $k=\{-\omega,\vec k\}$. Substituting this fluctuation in action (\ref{action}),
we get (\ref{fullaction})\footnote{$a_3,a_6$ are zero here}. Now, we rewrite this
action (\ref{action}) as equation of motion piece (which vanishes on-shell) and
boundary term \footnote{We ignore contribution from horizon as \cite{son4}}.
Thus, on-shell $S, S_{GH}, S_{CT}$  become,
\ben
S&=&\frac{1}{16 \pi G_5}\int_{r=\delta}\frac{d^4k}{(2 \pi)^4}L(\phi(r,k)), \nn \\
S_{GH}&=&\frac{1}{8 \pi G_5}\int_{r=\delta}\frac{d^4k}{(2 \pi)^4}L_{GH}(\phi(r,k)), \nn \\
S_{CT}&=&\frac{1}{16 \pi G_5}\int_{r=\delta}\frac{d^4k}{(2 \pi)^4}L_{CT}(\phi(r,k)),
\een
where we  define,
\ben
L(\phi(r,k))&=& a2(r) \phi'(r,k) \phi^*(r,k) \nn \\
&&+ {a4(r)-a5'(r) \over 2}\phi(r,k) \phi^*(r,k),\nn \\
L_{GH}(\phi(r,k))&=&2(g1) \phi(r,k) \phi^*(r,k) + 2(g2) \phi'(r,k) \phi^*(r,k),\nn \\
L_{CT}(\phi(r,k))&=&(  c_0 + c_1 \omega^2 + c_2 q^2) \phi(r,k) \phi^*(r,k).
\een
The coefficients $a_2,a_4,a_5$ are given in (\ref{boeff}) and other coefficients are,
\ben
g_1&=&\frac{2-r^2}{r^2}, \,\,\,\,\,\,\,\ g_2= -2\frac{1-r^2}{r} \nn \\
c_0&=&-3 \frac{\sqrt{1-r^2}}{r^2}, \,\,\,\,\,\ c_1=\frac{1}{4 r \sqrt{1-r^2}} \nn \\
c_2&=&- \frac{\sqrt{1-r^2}}{4 r}.
\een
The retarded Green's function is defined at asymptotic infinity as,
\be
G^R(k)=2\lim_{r \rightarrow 0}\frac{(L+L_{GH}+L_{CT})|_{on-shell}}{\phi_0(k)\phi_0(-k)}.
\ee
Here, $\phi_0(k)$ is the boundary value of the fluctuation (\ref{phifu}).
The retarded Green's function $G^R$ is a function of boundary momenta $k_{\mu}=(\omega,0,0,q)$.
Now the leading action and the Gibbons-Hawking action get divergences
from $\phi'(r,k) \phi^*(r,k)$ and $\phi(r)\phi^*(r,k)$ parts. Both these divergences
get canceled by the counter-term action which is always proportional to only
$\phi(r) \phi^*(r,k)$. In this case of leading Einstein's gravity, it is even
more simplified. Divergences coming from $\phi(r)\phi^*(r,k)$ piece of
leading and Gibbons-Hawking action gets canceled by momentum independent
piece of Counter-term action. It turns out that there is a cancellation among
the corresponding coefficients as,
\be
\lim_{r \rightarrow 0}({1\over 2}(a4(r)-a5'(r))+2 g1 + c_0)={1 \over 2},
\ee
i.e. the final contribution from $\phi(r)\phi^*(r,k)$ piece is only a finite number
${1 \over 2}$. As the graviton fluctuation $\phi(r,\omega, \vec k)= \phi_0(1+ F(r,\omega,\vec k))$
and moreover $\lim_{r \rightarrow 0}F(r,\omega,\vec k)=0$, we see that the
$\phi(r)\phi^*(r)$ term above only contribute to pressure (the $\omega$
independent piece of $G^R$). It would never contribute to any transport coefficient.

Also, the divergences coming from $\phi'(r,k) \phi^*(r,k)$ piece of original and
Gibbons-Hawking action, get canceled with the piece of the counter-term
proportional to $\omega^2, q^2$ ($c_1\,\,\ and \,\,\ c_2$ are purely
divergent at boundary). Here, the situation is more subtle, as there is no cancellation
among the coefficients. One actually needs to put the solution of
$\phi(r,\omega, \vec k)$ to see the cancellation.

The overall lesson from this detailed analysis is that counter-term only cancel the
UV divergences in usual holographic renormalization process and at most contribute
to pressure of the boundary plasma. It has no effects on any transport coefficients.
In \cite{Baier}, the author have computed second order transport coefficients for
the plasmas dual to leading Einstein's gravity following this usual approach. The
results are as follows,
\be\label{HOTC}
\tau_{\pi}={2- \ln 2 \over 2 \pi T}, \,\,\,\,\,\,\,\,\ \kappa={\eta \over \pi T}.
\ee
These results match with the one we obtained in (\ref{htf}) by solving the flow equations.
%%%%%%%%%%%%%%%%%%%%%%%%%%%%%%%%%%%%%%%%%%%%%%%%%%%%%%%%%%%%%%%%
%%%%%%%%%%%%%%%%%%%%%%%%%%%%%%%%%%%%%%%%%%%%%%%%%%%%%%%%%%%%%%%%

%%%%%%%%%%%%%%%%%%%%%%%%%%%%%%%%%%%%%%%%%%%%%%%%%%%%%%%%%%%%%%
%%%%%%%%%%%%%%%%%%%%%%%%%%%%%%%%%%%%%%%%%%%%%%%%%%%%%%%%%%%%%%%%

\section{Equivalence of Boundary Terms}\label{BOUN}

In this appendix we will show explicitly why the transport coefficients
computed from the original action and the effective action are same, even
for any higher derivative theory. It was already noticed \cite{BD1}, that the
two would give same first-order transport coefficient $\eta$ with a
suitable choice of the overall normalization constant. Here, we show that, not
just the first order transport coefficients, rather any higher order transport
coefficients computed from the original action and the effective action are same.

We consider a general class of action for $\phi$ which appears when the higher
derivative terms are made of different contraction of Ricci tensor, Riemann
tensor, Weyl tensor, Ricci scalar etc. or their different powers. Since, all
these tensors involve two derivatives of metric they can only have terms like
$\partial_a \partial_b \Phi(r,x)$ and its lower derivatives. Therefor the most generic
quadratic (in $\Phi(r,x)$,
in linear response theory) action for this kind of higher derivative gravity
has the following form (in
momentum space)\footnote{In all the expressions we have omitted $k$ dependence
  of $\phi$.}
\ben\label{fullaction}
S&=&\nt \intk dr \bigg[
 a1(r) \phi (r)^2+ a2(r)  \phi ' (r)^2 \nn \\
 && \hspace{0.5 cm} + a4(r) \phi(r)
   \phi'(r)
+ \alpha' \ a6(r) \phi''(r) \phi'(r) \nn \\
&& \hspace{0.5 cm} + \alpha' \ a3(r) \phi''(r)^2 + a5(r) \phi(r) \phi''(r)\bigg]
\een
where,
\ben\label{boeff}
a1(r)&=&
\frac {-8 r^2 + \omega ^2 r + 8} {4 r^3 - 4 r^5} + \alpha' \
f2(r) \nonumber \\
a2(r)&=&
-3 r + \frac {3} {r} + \alpha' \ h2 (r)  \nonumber \\
a4(r)&=&
- \frac {6} {r^2} - 2 + \alpha' \ g2 (r) \nonumber \\
a5(r)&=&
-4 r + \frac {4} {r}  +\alpha' \ j2 (r)
\een
and $a3(r), a6(r), j2(r), g2(r), h2(r)$ and $f2(r)$ depends on
higher derivative terms in the action and hence are computed purely from
the background solution with $\app \rightarrow 0$. Among these coefficients
$a3$ is special, as, it couples to $\phi''^2$. All four derivatives act on the
graviton fluctuation and thus $a3$ only depends on metric functions in (\ref{sol})
and there r-derivatives. It is easy to convince ourselves that
$a3 \propto r (r^2-1)^2 f(r, \app)$, where $f(r,\app)$ is a function that
depend on the higher derivative terms and finite (constant or 0) at the
boundary $r \rightarrow 0$.
Now let us write the effective Lagrangian as follows,
\ben
S_{eff}&=& {1 + \alpha'
\Gamma \over 16 \pi G_5} \intk dr \bigg[
\frac {4 r \left (r^2 - 1 \right)^2
\phi'(r)^2 - \omega^2 \phi(r)^2} {4 r^2 \left (r^2 - 1 \right)}
\nonumber \\
&&  + \alpha' \bigg (
b2
(r) \phi(r)^2 + \ b1 (r) \phi'(r)^2 \bigg )\bigg] \ .
\een
Demanding that the equation of motion (up to order $\alpha'$) of $\phi$ derived from
the original action and the above action are same we get,
\ben
b1(r)&=&{1 \over 2 r \left (r^2 - 1 \right)^2} [ (-4 r^3 - 12 r + \omega
^2) a3(r)
\nonumber \\
&& +  (r^2 - 1)
(2 \kappa
r^4 - a6'(r) r^3 - 4 \kappa r^2 + 2 a3'(r) r^2 \nonumber \\
 &&+ 2 (r^2 -
1 ) h2(r) r - 2  (r^2 - 1 ) j2(r) r +
a6' (r) r \nn \\
&& + 2 \kappa + 2 a3' (r))]
\een
\ben
b2(r)&=&
-{1\over 16 r^2 \left (r^2 -
      1 \right)^4} \bigg [ (\omega ^4 + 144 r^3 \omega^2 )
a3(r) \nonumber \\
&&  + 4  (r^2 - 1 ) \bigg (-4 r^2 f2(r) (r^2 - 1)^3  \nonumber \\
&& + ((\omega^2 \kappa - 2 r^2  (r^2 - 1 ) j2'' (r) ) (r^2 - 1) \nonumber \\
&& + 2 r^2 g2' (r)  (r^2 - 1)^2 + r \omega^2 a3'' (r) )  (r^2 -1 ) \nn \\
&&+  (1 - 11 r^2 )\omega^2 a3' (r)\bigg) \bigg]\ .
\een
The boundary terms coming from the original action (after adding {\it
Gibbons-Hawking} boundary terms) are given by
\footnote{There was a sign error in \cite{BD1}},
\ben
S^{{\cal B}}&=& \nt \intk\bigg[
-\frac {\phi (r)^2} {r^2} + \phi (r)^2 +
 r \phi ' (r) \phi (r) \nn \\
 &&- \frac {\phi ' (r)
        \phi (r)} {r}
+ \alpha' \bigg (\frac {1} {2} g2 (r) \phi (r)^2 - \frac {1} {2}
         j2' (r) \phi (r)^2 \nonumber \\
  &&+( h2 (r)  - j2 (r)
         - \frac {
a6' (r)}{2})\phi ' (r) \phi
        (r) \nonumber \\
  &&+\frac {
a3' (r) \left (\phi (r) \omega^2 +
          4 \left (r^4 - 1 \right)
               \phi ' (r) \right) \phi (r)} {4 r \left (r^2 -
          1 \right)^2} \nonumber \\
 &&- \frac {a3 (r)
           (6 r \phi (r)\phi ' (r) \omega^2)} {4 r
           \left (r^2 - 1 \right)^3} \nonumber \\
           &&- \frac {a3 (r)
           \left ( \left (r^2 -
              1 \right) \left (8 r^3 + 24 r -
\omega^2 \right) \phi ' (r) \right) \phi (r)} {4 r
           \left (r^2 - 1 \right)^3} \nonumber \\
  &&- \frac
{a3 (r) \phi ' (r) \left (\phi (r)
              \omega^2 +
         4 \left (r^4 - 1 \right) \phi ' (r) \right)} {4 r
           \left (r^2 - 1 \right)^2} \nonumber \\
  &&- a3 (r) \phi ' (r) \left (-\frac {\phi (r)
              \omega^2} {2 r \left (r^2 -
             1 \right)^2} - \frac {\left (r^4 - 1 \right) \phi
               ' (r)} {r \left (r^2 - 1 \right)^2} \right) \bigg )\bigg]\ .\nn \\
\een
And the boundary terms coming from the effective action are given by,
\ben
S_{seff}^{{\cal B}}&=& {1 \over 16 \pi G_5} \intk \bigg[
\left (r - \frac {1} {r}
\right) \phi (r) \phi ' (r) \nonumber \\
&&  + {\alpha' \over 2 r \left (r^2 - 1 \right)^2} \bigg (\ \phi (r) (2
            \Gamma  \left (r^2 -
           1 \right)^3
+ (- a6' (r) r^3
\nonumber \\
&&  + 2 a3' (r) r^2 +
           2 \left (r^2 - 1 \right) h2 (r) r - 2
                \left (r^2 - 1 \right) j2 (r) r
\nonumber \\
&&  + a6' (r) r  + 2
               a3' (r) ) \left (r^2 -
          1 \right) \nn \\
          &&  + \left (-4 r^3 - 12 r +
\omega^2 \right) a3 (r)
) \phi ' (r)\bigg ) \bigg]\ .
\een
 Now, it is interesting to compute the difference between these two boundary terms and the result
is\footnote{It has been shown in \cite{BD1} that $\Gamma=0$.},
\ben\label{diff}
S^{\cal B}- S^{\cal B}_{eff}&=& \nt \intk\bigg[
-\frac {\phi (r)^2} {r^2} + \phi (r)^2  \nonumber \\
&& \,\,\,\,\,\ +
 \alpha' \bigg (\frac {1} {2} g2 (r) \phi (r)^2 - \frac {1} {2}
         j2' (r) \phi (r)^2 \nonumber \\
&& \,\,\,\,\,\ +\frac {
a3' (r) \omega^2 \phi (r)^2} {4 r \left (r^2 -
          1 \right)}  - \frac {a3 (r)
           \left (6 r \omega^2 \right) \phi (r)^2} {4 r
           \left (r^2 - 1 \right)^3}\bigg ) \nonumber \\
\een
The term proportional to $a3$ in the parenthesis of (\ref{diff}) vanishes
at the boundary whereas the term proportional to $a3'$ gives a pure UV divergent
piece and a vanishing piece, due to the property of $a3$ mentioned above. This
is true irrespective of the choice of the higher derivative terms. Thus, we see
that the two boundary terms differ only by terms which are either purely divergent
or of the form $g(r)\phi^2$, where $g(r)$ is any function of $r$. The divergent terms would get canceled once appropriate
boundary terms are added (which has been discussed in sections \ref{model} and \ref{modelhd}). The terms
proportional to $\phi^2$ can only contribute to pressure of the boundary theory
and are not important for the computation of transport coefficients of the boundary
plasma. Thus we see that, it is obvious that the transport coefficients coming from
the original action and the boundary action are same.

Here we have considered only $R^{(n)}$ gravity theory. A more rigorous proof is required
for theories involving covariant derivatives of curvature tensors and scalars.

%%%%%%%%%%%%%%%%%%%%%%%%%%%%%%%%%%%%%%%%%%%%%%%%%%%%%%%%%%%%%%%%
%%%%%%%%%%%%%%%%%%%%%%%%%%%%%%%%%%%%%%%%%%%%%%%%%%%%%%%%%%%%%%%%

\section{Functions appeared in String theory corrected action}\label{weylapp}

\ben
{\cal A}_0^{W^4}(r,k)&=& -\frac{r_0^2 (q^2 (r^2-1)+\omega ^2)}{4
   r^2 \left(r^2-1\right)}\nn\\
&& -{1\over 4 \left(r^2-1\right)}\bigg(\gamma  (4 q^2 r^2
   ((245 r^4 \nn\\
&&-407 r^2+162) r_0^2+10 r \omega
   ^2)\nn\\
&&+3 \left(221 r^4-191 r^2+25) \omega ^2
   r_0^2\right)\bigg)
\een
and
\ben\label{a1w}
{\cal A}_1^{W^4}(r,k)&=& \frac{\left(r^2-1\right) r_0^4}{r}\nn \\
 &&+ 3 r \left(r^2-1\right) \gamma  r_0^2 \bigg((43 r^4 \nn \\
 &&+47
   r^2-25) r_0^2
    +16 q^2r^3\bigg).
\een

%%%%%%%%%%%%%%%%%%%%%%%%%%%%%%%%%%%%%%%%%%%%%%%%%%%%%%%%%%%%%
%%%%%%%%%%%%%%%%%%%%%%%%%%%%%%%%%%%%%%%%%%%%%%%%%%%%%%%%%%$%$

\section{Functions appeared in four derivative action}\label{gbapp}

\ben
{\cal A}_0^{\rm GB}(r,k)&=& -\frac{q^2
   \left(r^2-1\right)+\omega ^2}{4 r^2 \left(r^2-1\right)} \nn\\
&&+ {\alpha ' \over 12 r^2 \left(r^2-1\right)} \bigg(q^2 (2 \beta _3 (13 r^2-3 r
   \omega ^2-13)\nn\\
&& +130 (r^2-1) \beta _1+(-36
   r^4+25 r^2+11) \beta _2)\nn\\
&&+\omega ^2 ((6
   r^2-11) \beta _2+130 \beta _1+26 \beta
   _3)\bigg)
\een
and
\ben\label{a1gb}
{\cal A}_1^{\rm GB}(r,k)&=& r-\frac{1}{r}\nn\\
&-&\frac{\left(r^2-1\right) \left(\left(18 r^2-13\right) \beta
   _2+110 \beta _1+22 \beta _3\right) \alpha '}{3
   r} .\nn \\
\een

%%%%%%%%%%%%%%%%%%%%%%%%%%%%%%%%%%%%%%%%%%%%%%%%%%%%%%%%%%%%%%%%%%%%%%%%
%%%%%%%%%%%%%%%%%%%%%%%%%%%%%%%%%%%%%%%%%%%%%%%%%%%%%%%%%%%%%%%%%%%%%%%%

\section{Functions appeared in Higher-derivative Charged black-hole action}\label{hdcb}

\ben
{\cal A}_1^{CB}&=&\frac{r^5(\omega ^2-q^2)+q^2 r_0^4 r}{2 (r^4-r_0^4)}\nn \\
&+&\frac{c_1 \alpha' (11 r^8 (\omega ^2-q^2)-r_0^4 r^4(25 q^2+6 \omega ^2)+36 q^2 r_0^8)}{6 r^3
   (r^4-r_0^4)} \nn \\
&+& \frac{Q^2}{2 r_0^2 (r^2-r_0^2)
   (r^2+r_0^2)^2} \bigg[r^3 \omega ^2+\frac{\alpha'}{3 r^5} \bigg(c_1 [28 q^2 r_0^6 r^2 \nn \\
   &-&r^8 (36 q^2+127 \omega ^2)-4 r_0^2 r^6 (7 q^2-3 \omega ^2)-8 q^2 r_0^8\nn \\
   &+&4 r_0^4 r^4 (11 q^2+3
   \omega ^2)]-24 c_2 [r_0^2 r^6 (2 q^2-3 \omega ^2)-2 q^2 r_0^6
   r^2 \nn \\
   &+&r_0^4 r^4 (2 q^2-3 \omega ^2)-2 q^2 r_0^8+4 r^8 \omega ^2]\bigg) \bigg] +  {\cal O}(Q^4)
\een
and
\ben
{\cal A}_1^{CB}&=&\frac{1}{2} \left(r r_0^4-r^5\right)
-\frac{c_1 \alpha' \left(13 r^4-18 r_0^4\right) \left(r^4-r_0^4\right)}{6 r^3}\nn \\
 &+&   \frac{Q^2 (r^2-r_0^2)}{2 r r_0^2 } \bigg[1-\frac{\alpha'}{3
   r^4} \bigg(24 c_2 (4 r^4-3 r_0^2 r^2-3 r_0^4) \nn \\
   &&+ c_1 (101 r^4-156 r_0^2 r^2-120 r_0^4)\bigg)\bigg] + {\cal O}(Q^4). \nn \\
\een

%%%%%%%%%%%%%%%%%%%%%%%%%%%%%%%%%%%%%%%%%%%%%%%%%%%%%%%%%%%%%%%%%%%%%%%%%%

\end{document}